\documentstyle[12pt,aasms4]{article}

\def\MBH{{M_{\rm BH}}}
\def\Mdot{{\dot M}}
\def\Mdotc{{\dot M_{\rm c}}}
\def\Ms{{M_\odot}}
\def\OmegaK{{\Omega_{\rm K}}}

\def\rg{{r_{\rm g}}}

\received{4 April 1997}
\accepted{6 June 1997}

\lefthead{MANMOTO, MINESHIGE & KUSUNOSE}
\righthead{SPECTRUM OF ADVECTION DOMINATED ACCRETION FLOW}

\begin{document}

\baselineskip 5mm

\title{SPECTRUM OF OPTICALLY THIN\\
    ADVECTION-DOMINATED ACCRETION FLOW AROUND A BLACK HOLE:\\
    APPLICATION TO SAGITTARIUS A*}
    
\author{T. Manmoto and S. Mineshige}
\affil{Department of Astronomy, Faculty of Science,
    Kyoto University, Sakyo-ku, Kyoto 606-01, Japan}
    
\and

\author{M. Kusunose}
\affil{Department of Physics, School of Science
    Kwansei Gakuin Univeristy, 1-155 Uegahara Ichibancho
    Nishinomiya 662, Japan}
    
\begin{abstract}
The global structure of optically thin advection dominated accretion flows 
which are composed of two-temperature plasma around black holes is calculated.  
We adopt the full set of basic equations including 
the advective energy transport in the energy equation for 
the electrons.  The spectra emitted by the optically thin accretion flows 
are also investigated. The radiation mechanisms which are taken into accout 
are bremsstrahlung, synchrotron emission, and Comptonization. 
The calculation of the spectra and that of the structure 
of the accretion flows are made to be completely consistent by calculating 
the radiative cooling rate at each radius.  
As a result of the advection domination for the 
ions, the heat transport from the ions to the electrons becomes practically 
zero and the radiative cooling balances with the advective {\it heating} 
in the energy equation of 
the electrons.  Following up on the successful work of Narayan et al.  
(1995), we applied our model to the spectrum of Sgr A*.  We find that the 
spectrum of Sgr A* is explained by the optically thin advection dominated 
accretion flow around a black hole of the mass $\MBH=10^{6}\Ms$.  The 
parameter dependence of the spectrum and the structure of the accretion 
flows is also discussed.
\end{abstract}

    
\section{Introduction}

Ever since the pioneering studies of steady thin accretion disks by Shakura \& 
Sunyaev (1973, hereafter SS),
the model of thin accretion disks has been applied successfully to 
low energy emission from astrophysical objects, such as 
dwarf novie and pre-main-sequence stars.  
However, the model has been less successful in
modeling of high energy emission from Galactic black hole candidates and active 
galactic nuclei which are considered to be powered by accreting black 
holes. The main problem lies in 
the fact that the thin accretion disks can not reproduce 
the observed spectra of such systems.
The thin accretion disk model which SS originally proposed 
assumes that the disk is optically thick in the vertical direction
and radiates the energy generated by the viscosity locally. 
This model predicts that the generated spectrum is 
multi-colored black body, which cannot explain the observed power-law 
spectra of X-rays from AGNs
and Galactic black hole candidates
even though it can explain the UV bump of the AGN or the soft 
state spectrum of the Galactic black hole candidates.
More fundamentally, 
although it is generally believed that QSOs and Seyfert galaxies are powered by the gas 
accretion onto a super-massive black hole of the mass $\MBH\sim 10^8\Ms$, 
the thin accretion disks onto such super-massive 
black holes are too cool to generate high energy photons 
which are observed in many QSOs and Sayfert galaxies.

The model which is investigated by Shapiro, Lightman \& Eardley (1976, 
hereafter SLE) 
is quite attractive in that the accreted gas is optically thin and is much 
hotter than that in the SS solution 
and is hot enough to produce high energy photons.  
SLE considered two-temperature plasma with ions being much 
hotter than electrons, which enabled quantitative studies of the 
non-blackbody spectra.  
SLE model has been applied to explain the spectrum of X-ray binaries 
and active galactic nuclei successfully.
However, it is known that the optically thin hot accretion disk 
which is considered in SLE is thermally unstable (Piran 1978).
If the accretion disk is heated up, 
then the disk expands and the density decreases, so that the
bremsstrahlung cooling rate decreases. The reduced cooling then causes 
the gas to become even hotter, leading to a runaway thermal instability.
For this reason it is not likely that such hot accretion disks exist in real 
systems for much longer than the thermal timescale.

The models introduced so far are local solutions in the sense that the heat 
generated via viscosity is locally radiated away efficiently, which 
corresponds to neglecting the advective energy transport
in the energy equations.  When 
plasma cannot emit radiation efficiently, the heat generated via viscosity 
is advected inwards as the internal energy of the plasma.
Abramowicz et al.  (1988) investigated the effect of advection term in 
their ``slim disk'' model in the optically thick case,
and made it clear 
that there exists advection dominated branch where the viscous heating is 
balanced with the advection term rather than the radiative cooling term at
high mass accretion rates.  

The optically thin advection dominated solution at low mass accretion 
rates is studied intensively by 
Abramowicz et al. (1995) and Narayan \&Yi  
(1995) (see also Ichimaru 1977, Matsumoto et al. 1985).  
Although they claimed that optically thin advection dominated solution is 
thermally stable for the long wavelength perturbations, Kato et 
al.  (1996) showed the possibilities of the instability against the short 
wavelength perturbations. Manmoto et al.  (1996) demonstrated that such an 
instability is favored for explaining rapid X-ray fluctuations from 
Galactic black hole candidates and does not affect the global stability of 
the accretion flows.  

The optically thin advection dominated accretion flows are applied to 
explain the observed spectra of accreting black holes.
Narayan \& Yi (1995) investigated the self-similar solution 
which is used later to calculate the spectrum from several 
low-luminosity accreting systems with great success.
As a next step, 
the calculations of full global steady solutions were awaited
for further investigations. 
Chen et al.  (1997) solved optically thin 
advection dominated solution globally, but their solutions are those of  
one-temperature plasma and do not include detailed radiation mechanisms.
Narayan et al.  (1997) derived global solution for optically thin advection 
dominated accretion flows and showed that the self-similar solution is 
a good approximation at the radius far enough from the outer and the inner 
boundaries. This means that the spectra which are derived by using the 
self-similar solution
may be modified, because considerable amount of photons may come from the 
hot region near the inner boundary.  Narayan et al.  (1997)
did calculate the spectra from two-temperature accretion flows 
with their global solutions, but they
treated the electron energy equations locally, 
neglecting the effect of the electron advection.
Nakamura et al.  (1996) were the first to solve the energy equations for ions and electrons 
and obtained the global two-temperature advection dominated solutions,
and showed that the temperature profiles which are crucial to the 
generated spectra are largely modified when 
the effect of electron advection is taken into account.
However, Nakamura et al.  (1996) focused their attentions on the structure of the optically thin 
accretion disks and did not investigate the spectra from the disks.

The study on the spectrum from optically thin advection dominated 
accretion flows with full global treatment of the basic equations is yet to be done.  
Thus we are motivated to consistently solve full set of equations including 
the energy equation for ions and electrons with detailed radiation 
mechanisms and obtain the spectra from the optically thin accretion flows.

In section 2, we present the physical assumptions and the basic equations 
of our model.  We show the results of our 
calculations in section 3.  We 
then apply our model to Sgr A* (the central core of our Galaxy) in section 
4.  We conclude in section 5 with a summary and discussion.

\section{Accretion Flow Model}

\subsection{Physical Assumptions}

We consider an optically thin, steady axisymmetric accretion flow around 
a black hole.
To investigate the spectra generated by the optically thin gas flows, we 
discuss gas dynamics in the context of two-temperature plasma.
Assuming the existence of randomly oriented magnetic fields which possibly 
originate from the turbulence in the gas flow, we take total pressure 
$p$ to be 
\begin{equation}
	p=p_{\rm gas}+p_{\rm mag},
	\label{eq2.1}
\end{equation}
where $p_{\rm gas}$ is the gas pressure, $p_{\rm mag}$ is the magnetic 
pressure.
We neglect the radiation pressure in this paper because the optically thin accretion 
flows we consider is always gas pressure dominated.
We take the ratio of 
gas pressure to the total
pressure as a global parameter which we designate as $\beta$.  Technically we 
need to solve the magnetic field self-consistently with the gas dynamics, 
but it is beyond the scope of this paper to treat full 
magneto-hydrodynamics equations.
Due to the two-temperature assumption, we write $p_{\rm gas}$ as
\begin{equation}
	p_{\rm gas}=\beta p=p_{i}+p_{e}={\rho \over {\mu _{i}}}{k 
	\over {m_{\rm H}}}T_{i}+{\rho 
	\over {\mu _{e}}}{k \over {m_{\rm H}}}T_{e}.
	\label{eq2.2}
\end{equation}
Here and hereafter subscripts $i$ and $e$ indicate the quantities for 
ions and electrons, respectively. In eq. (\ref{eq2.2}), $T$ is the 
temperature, $\rho$ is the density, and $\mu$ is the mean molecular weight 
which is given by
\begin{equation}
	\mu _{i}=1.23,\quad\mu _{e}=1.14,
	\label{eq2.3}
\end{equation}
where numerical values correspond to the cosmic abundance.
We estimate magnetic field $B$ via magnetic pressure by the following equation:
\begin{equation}
	p_{\rm mag}=\left( {1-\beta } \right)p={{B^2} \over {8\pi }}.
	\label{eq2.4}
\end{equation}

\subsection{Basic Equations}

We adopt cylindrical coordinate system ($r$,$\varphi$,$z$) to describe 
axisymmetric (
${\partial  \over {\partial \varphi }}=0$
) accretion flows.  Basic equations which describe the 
dynamics of the accretion disks are those of mass conservation, Euler 
equations which comprises three spatial components, and the energy 
equations.
Mass conservation gives
\begin{equation}
	{\partial \over {\partial r}}\left( {r\rho v_{r}} \right)+r{\partial \over 
	{\partial z}}\left( {\rho v_{z}} \right)=0,
	\label{eq2.5}
\end{equation}
where $v_{r}$ and $v_{z}$ are the 
radial and the vertical velocity respectively.
The radial component of Euler equation is 
\begin{equation}
	{\partial \over {\partial r}}\left( {r\rho v_{r}^2} \right)+r{\partial 
	\over {\partial z}}\left( {\rho v_{r}v_{z}} \right)=-\rho \left( {r{{\partial 
	\psi } \over {\partial r}}-v_{\varphi} ^2} \right)-r{{\partial p} \over 
	{\partial r}},
	\label{eq2.6}
\end{equation}
where $v_{\varphi}$ is the azimuthal 
velocity and $\psi$ is the potential energy.  To simulate general relativistic 
effects, we adopt pseudo-Newtonian potential 
\begin{equation}
	\psi =-G\MBH/ (R-\rg)
	\label{eq2.7}
\end{equation}
with $\MBH$ being the mass of the black hole, $\rg$ the 
Schwarzschild radius, and $R=(r^{2}+z^{2})^{1/2}$ the distance from 
the central black hole.
This potential is known to 
represent the dynamical aspects of general relativistic effects quite well for $r>2\rg$
(Paczy\'nski \& Wiita 1980) and greatly simplifies the basic equations.  
Technically we need to solve the basic equations in the Schwarzschild 
or Kerr metric especially for the photon propagation. However solving 
the equations including the photon propagation in the full 
relativisitc metric is out of the scope of current paper.
Azimuthal component of Euler equation is the conservation of angular momentum 
which gives
\begin{equation}
	{\partial  \over {\partial r}}\left( {r^2\rho v_rv_\varphi} \right)+
	r{\partial  \over {\partial z}}\left( {r\rho v_\varphi v_z} \right)=
	{\partial  \over {\partial r}}\left( {r^2\tau _{r\varphi }} \right).
	\label{eq2.8}
\end{equation}
Here $\tau_{r \varphi}$ is the $r \varphi$-component of the stress tensor. 
According to the conventional $alpha$-prescription of sheer viscosity, 
$\tau_{r \varphi}$ can be written as
\begin{equation}
	\tau _{r\varphi }=\alpha p{{d\ln \Omega } \over {d\ln r}}
	{\Omega  \over {\Omega _k}},
	\label{eq2.8.5}
\end{equation}
where $\Omega$ is the angular velocity and $\OmegaK$ is the Keplerian angular velocity 
on the equatorial plane.
In our paper, we take $\tau_{r \varphi}$ to be simply proportional to the local pressure 
$p$:
\begin{equation}
	\tau _{r\varphi }=-\alpha p,
	\label{eq2.9}
\end{equation}
where $\alpha$ is the dimensionless viscosity parameter which in 
general is 
considered to be around $0.01\sim0.1$.
Two-temperature assumption requires two energy equations i.e. one for ions and 
one for electrons, both of which contain advective energy transport terms.  The energy 
equation for each species is 
\begin{equation}
	\rho T_i\left( {v_r{{\partial s_i} \over {\partial r}}+v_z{{\partial s_i} 
	\over {\partial z}}} \right)=\tau _{r\varphi }r{{\partial \Omega } \over 
	{\partial r}}-\lambda _{ie},
	\label{eq2.10}
\end{equation}
\begin{equation}
	\rho T_e\left( {v_r{{\partial s_e} \over {\partial r}}+v_z{{\partial s_e} 
	\over {\partial z}}} \right)=\lambda _{ie}-q_{\rm rad}^-,
	\label{eq2.11}
\end{equation}
where $s$ is the specific entropy, 
$\lambda_{ie}$ is the volume energy transfer rate from ions to electrons,
and $q_{\rm rad}^{-}$ is the volume radiative cooling rate.
We assume that the energy is transferred 
from ions to electrons via Coulomb collisions only.  Stepney \& Guilbert 
(1983) give an explicit expression:
\begin{equation}
    \begin{array}{l}
    \lambda _{ie}=1.25\times {3 \over 2}{{m_e} \over 
	{m_p}}n_en_i\sigma _{\rm T}c{{\left( {kT_i-kT_e} \right)} \over {K_2\left( {1/ 
	\theta _e} \right)K_2\left( {1/ \theta _i} \right)}}\ln \Lambda\\
	\quad\times \left[ {{{2\left( {\theta _e+\theta _i} \right)^2+1} \over 
	{\left( {\theta _e+\theta _i} \right)}}K_1\left( {{{\theta _e+\theta _i} 
	\over {\theta _e\theta _i}}} \right)+2K_0\left( {{{\theta _e+\theta _i} 
	\over {\theta _e\theta _i}}} \right)} \right]
	\end{array}
	\label{eq2.12}
\end{equation}
where the $K$'s are modified Bessel functions, $\ln \Lambda=20$ is the 
Coulomb logarithm, and 
$\theta \equiv kT/ m_ec^2$
is the dimensionless temperature.
Note that the energy 
equations assume that the ions and the electrons are in thermal equilibrium by 
some mechanisms.  The thermodynamic relations are 
\begin{equation}
	\rho T_ids_i={1 \over {\gamma -1}}\left( {dp_i-\gamma {{p_i} \over \rho 
	}d\rho } \right),
	\label{eq2.13}
\end{equation}
\begin{equation}
	\rho T_eds_e={1 \over {\gamma -1}}\left( {dp_e-\gamma {{p_e} \over \rho 
	}d\rho } \right),
	\label{eq2.14}
\end{equation}
where $\gamma=5/3$ is the adiabatic index.
In the energy equations we have assumed viscous heating acts only on ions 
because an ion particle is much heavier than an electron particle.

Now we have full set of basic equations which describe the steady 
accretion flow around a black hole.
However, it is rather a difficult problem to solve above partial differential 
equations with respect to $r$ and $z$ plus radiative transfer equations to 
obtain global steady solutions.  To lessen the number of variables, 
we fix the vertical structure before we solve the full spatial 
equations, assuming the accretion flow is geometrically thin.
The validity of the assumption that the disk is 
geometrically thin will be discussed later.
As a first order approximation, we adopt isothermal structure in the 
vertical direction, which means that the sound velocity 
$c_{\rm s}\equiv \left( {p/ \rho } \right)^{1/ 2}$
is independent 
of $z$.  Furthermore, we also assume radial and azimuthal velocities are 
independent of $z$.
The vertical structure is obtained analytically by solving the remaining 
vertical component of Euler equation which represents hydrostatic balance 
in the vertical direction:
\begin{equation}
	{{\partial p} \over {\partial z}}=-\rho {{\partial \psi } \over {\partial 
	z}}=-\rho \OmegaK^2z.
	\label{eq2.15}
\end{equation} 
The last equation of eq. (\ref{eq2.15}) assumes that the disk is geometrically 
thin.
The density distribution in the vertical 
direction is 
\begin{equation}
	\rho \left( {r,z} \right)=\rho \left( {r,0} \right)\exp \left( {-{{z^2} 
	\over {2H^2}}} \right).
	\label{eq2.16}
\end{equation}
Here, H is the vertical scale height defined by
\begin{equation}
	H\equiv c_s/ \Omega _{\rm K}.
	\label{eq2.17}
\end{equation}
With the vertical structure given above, we can now integrate above basic 
equations in the vertical direction and rewrite them as follows.
Mass conservation now gives
\begin{equation}
	\dot M=2\pi r\Sigma v_r,
	\label{eq2.18}
\end{equation}
where 
$\Sigma \left( r \right)\equiv\int_{-\infty }^\infty {\rho \left( {r,z} 
\right)}dz=2\sqrt {\pi / 2}\rho \left( {r,0} \right)H$
is the surface density. We have taken $v_{r}$ to be positive for the inward flow so that 
the mass accretion rate $\dot M$ takes positive value.
The height-integrated version of the radial component of eq. (\ref{eq2.6}) is
\begin{equation}
	v_r{{dv_r} \over {dr}}+{1 \over \Sigma }{{dW} \over {dr}}=r\left( {\Omega 
	^2-\OmegaK^2} \right)-{W \over \Sigma }{{d\ln \OmegaK} \over 
	{dr}},
	\label{eq2.19}
\end{equation}
where 
$W\left( r \right)\equiv\int_{-\infty }^\infty {p\left( {r,z} \right)}dz=2\sqrt 
{\pi / 2}p\left( {r,0} \right)H$
is the height-integrated pressure. The last 
term of the right hand side of eq. (\ref{eq2.19}) corresponds 
to the correction for the decrease 
of the radial component of gravitational force away from the equatorial 
plane (Matsumoto et al. 1984).
We can integrate the conservation of angular momentum (eq. 
[\ref{eq2.8}]) with respect to $r$ as 
well as $z$ and obtain following simple equation:
\begin{equation}
	\dot M\left( {l-l_{\rm in}} \right)=2\pi r^2\alpha W,
	\label{eq2.20}
\end{equation}
where $l_{\rm in}$ is the 
specific angular momentum swallowed by the central black hole.
The energy equations (eq. [\ref{eq2.10}], [\ref{eq2.11}]) are vertically integrated 
together with the thermodynamic relations (eq. [\ref{eq2.13}], 
[\ref{eq2.14}]) and now can 
be written as
\begin{equation}
	\dot M{{W_i} \over \Sigma }\left( {{{\gamma +1} \over {2\left( {\gamma 
	-1} \right)}}{{d\ln W_i} \over {dr}}-{{3\gamma -1} \over {2\left( {\gamma 
	-1} \right)}}{{d\ln \Sigma } \over {dr}}-{{d\ln \Omega _k} \over {dr}}} 
	\right)=\dot M\left( {l-l_{\rm in}} \right){{d\Omega } \over {dr}}+2\pi 
	r\Lambda _{ie},
	\label{eq2.21}
\end{equation}
\begin{equation}
	\dot M{{W_e} \over \Sigma }\left( {{{\gamma +1} \over {2\left( {\gamma 
	-1} \right)}}{{d\ln W_e} \over {dr}}-{{3\gamma -1} \over {2\left( {\gamma 
	-1} \right)}}{{d\ln \Sigma } \over {dr}}-{{d\ln \Omega _k} \over {dr}}} 
	\right)=2\pi r\left( {Q_{\rm rad}^--\Lambda _{ie}} \right),
	\label{eq2.22}
\end{equation}
where $\Lambda _{ie}\equiv\int_{-\infty }^\infty  {\lambda _{ie}
\left( z \right)}dz=\sqrt \pi H\lambda _{ie}\left( 0 \right)$
is the energy transfer rate from ions to 
electrons per unit surface area. $Q_{\rm rad}^-$ is the 
radiative cooling rate per unit surface area which will be explicitly 
given in the next subsection.
The energy equations can be written compactly as follows:
\begin{equation}
	Q_{{\rm adv},i}^-=Q_{\rm vis}^+-\Lambda _{ie},
	\label{eq2.22.5}
\end{equation}
\begin{equation}
	Q_{{\rm adv},e}^-=\Lambda _{ie}-Q_{\rm rad}^-.
	\label{eq2.23}
\end{equation}
\subsection{Radiation Mechanism}

Following the work of Narayan \& Yi (1995), we consider three processes for 
radiative cooling: bremsstrahlung, synchrotron radiation, and Comptonization of soft 
photons.
In order to obtain the spectrum generated by the accretion flow and the 
global structure of the accretion disk, we need to solve global radiation 
transfer problem in the radial and the vertical direction which involves 
self-absorption and incoherent scatterings.  We treat such complicated 
problem in a rather simplified way: 1) we assume locally plane parallel gas 
configuration at each radius and 2) we separate the Compton scattering process 
from the rest of radiation processes, i.e.  emission and absorption. This 
will not make serious error because the gas is so tenuous that generated 
photons rarely experience multiple scatterings before they escape.  Note 
that at such low frequencies as radio frequencies, 
the effect of free-free and synchrotron self-absorption 
is so large that the disk can not be considered optically thin, while the 
optical depth for scatterings is constant at all frequencies.
We first estimate the spectrum of unscattered photons at given radius by 
solving radiative diffusion equation in the vertical direction.  For the 
isothermal plane parallel gas atmosphere where density configuration is 
given by eq. (\ref{eq2.16}) the radiative diffusion equation can be solved and gives 
the energy flux $F_{\nu}$ of the unscattered photons at 
given radius (see Appendix B):
\begin{equation}
	F_\nu ={{2\pi } \over {\sqrt 3}}B_\nu \left[ {1-\exp \left( {-2\sqrt 
	3\tau _\nu ^*} \right)} \right],
	\label{eq2.24}
\end{equation}
where $\tau _\nu ^*\equiv{{\sqrt \pi } \over 2}\kappa _\nu (0)H$ is the optical depth for 
absorption of the accretion flow in the vertical direction 
with $\kappa_{\nu}(0)$ being the absorption coefficient on the 
equatorial plane.  Assuming LTE, we can write
$\kappa _\nu =\chi _\nu / \left( {4\pi B_\nu } \right)$
where $\chi _\nu =\chi _{\nu ,{\rm brems}}+\chi _{\nu ,{\rm synch}}$
is the emissivity. Note 
that equation (\ref{eq2.24}) includes the effects of free-free absorption and 
synchrotron self-absorption at low frequencies.

\subsubsection{Bremsstrahlung Emission}

In our model, electron temperature exceeds rest mass energy of an electron in 
some cases where electron-electron bremsstrahlung is important as well as 
electron-proton bremsstrahlung which dominates electron-electron 
bremsstrahlung in the classical temperature regime.  Thus the
cooling rate per unit volume for bremsstrahlung is
\begin{equation}
	q_{\rm br}^-=q_{ei}^-+q_{ee}^-.
	\label{eq2.25}
\end{equation}
Following Narayan \& Yi (1995), we employ
\begin{equation}
	q_{ei}^-=1.25n_e^2\sigma _{\rm T}c\alpha _fm_ec^2F_{ei}
	\left( {\theta _e} \right),
	\label{eq2.26}
\end{equation}
where $\alpha_{f}$ is the fine-structure constant and $F_{ei}$ is 
given by
\begin{equation}
	F_{ei}\left( {\theta _e} \right)=4\left( {{{2\theta _e} \over {\pi ^3}}} 
	\right)^{1/ 2}\left( {1+1.781\theta _e^{1.34}} \right)\quad for\;\theta 
	_e<1,
	\label{eq2.26.5}
\end{equation}
\begin{equation}
	F_{ei}\left( {\theta _e} \right)={{9\theta _e} \over {2\pi }}\left[ {\ln 
	\left( {1.123\theta _e+0.48} \right)+1.5} \right]\quad for\;\theta _e>1,
	\label{eq2.27}
\end{equation}
and
\begin{equation}
	\begin{array}{l}
	q_{ee}^-=n_e^2cr_e^2\alpha _fc^2{{20} \over {9\pi 
	^{1/ 2}}}\left( {44-3\pi ^2} \right)\theta _e^{3/ 2}\\
	\quad\times 
	\left( {1+1.1\theta _e+\theta _e^2-1.25\theta _e^{5/ 2}} 
	\right)\quad for\;\theta _e<1,
	\end{array}
	\label{eq2.27.5}
\end{equation}
\begin{equation}
	q_{ee}^-=n_e^2cr_e^2\alpha _fc^224\theta _e\left( {\ln 1.1232 \theta 
	_e+1.28} \right)\quad for\;\theta _e>1.
	\label{eq2.28}
\end{equation}
With the cooling rate given above, we can write the emissivity 
$\chi _{\nu ,{\rm brems}}$ as 
\begin{equation}
	\chi _{\nu ,{\rm brems}}=q_{\rm br}^-\bar G \exp \left( {{{h\nu } 
	\over {kT_e}}} \right),
	\label{eq2.29}
\end{equation}
where $\bar G$ is the Gaunt factor which is given by (see Rybicki \& 
Lightman 1979)
\begin{equation}
	\bar G={h \over {kT_{e}}}\left( {{3 \over \pi }{{kT_{e}} \over {h\nu }}} \right)^
	{1/ 2}\quad for\;{{kT_{e}} \over {h\nu }}<1,
	\label{eq2.29.5}
\end{equation}
\begin{equation}
	\bar G={h \over {kT_{e}}}{{\sqrt 3} \over \pi }\ln \left( {{4 \over \zeta }{{kT_{e}}
	\over {h\nu }}} \right)\quad for\;{{kT_{e}} \over {h\nu }}>1.
	\label{eq2.30}
\end{equation}
\subsubsection{Synchrotron Emission}

For the relativistic temperatures for the electrons, and in the presence of 
magnetic field which is of the same order as equipartition magnetic field, 
synchrotron emission is also very important. 
Following Narayan \& Yi (1995), the optically thin 
synchrotron emissivity by a relativistic Maxwellian distribution of 
electrons is 
\begin{equation}
	\chi _{\nu ,{\rm synch}}=4.43\times 10^{-30}{{4\pi n_e\nu } \over {K_2\left( 
	{1/ \theta _e} \right)}}I'\left( {{{4\pi m_ec\nu } \over {3eB\theta 
	_e^2}}} \right),
	\label{eq2.31}
\end{equation}
where $I'(x)$ is given by
\begin{equation}
	I'\left( x \right)={{4.0505} \over {x^{1/ 6}}}\left( {1+{{0.4} \over 
	{x^{1/ 4}}}+{{0.5316} \over {x^{1/ 2}}}} \right)\exp \left( 
	{-1.8899x^{1/ 3}} \right).
	\label{eq2.32}
\end{equation}
\subsubsection{Compton Scattering}

We proceed to consider the effect of Compton scattering.  We make use 
of the idea of energy enhancement factor which is derived by Dermer, Liang, \& 
Canfield (1991) and modified in 
part by Esin et al. (1996). 
The energy enhancement factor $\eta$ is defined as the 
average energy boost of a photon.  The prescription for $\eta$ is 
\begin{equation}
	\eta =\exp \left( {s\left( {A-1} \right)} \right)\left[ {1-P\left( 
	{j_{\rm m}+1,As} \right)} \right]+\eta _{\rm max}P
	\left( {j_{\rm m}+1,s} \right),
	\label{eq2.33}
\end{equation}
where $P$ is the incomplete gamma function and 
\begin{equation}
	\begin{array}{l}
	A=1+4\theta _e+16\theta _e^2,\\
	s=\tau _{\rm es}+\tau _{\rm es}^2,\\
	\eta _{\rm max}={{3kT_e} \over {h\nu }},\\
	j_{\rm m}={{\ln \left( {\eta _{\rm max}} \right)} 
	\over {\ln \left( A \right)}}.
	\end{array}
	\label{eq2.34}
\end{equation}
$\tau_{es}$ is the optical depth for scattering:
\begin{equation}
	\tau _{\rm es}=2n_e\sigma _{\rm T}H\times \max \left( {1,{1\over\tau 
	_{\rm eff}}} \right),
	\label{eq2.35}
\end{equation}
where $\tau _{\rm eff}\equiv\tau _\nu \left( {1+n_e\sigma _{\rm T}/ \kappa _\nu } \right)^{1/ 2}$
is the effective optical depth (Rybicki \& Lightman 1979).
Equation (\ref{eq2.32}) gives the correct estimate for the optical depth for 
scattering in the presence of the absorption.
Note that the simple treatment given above assumes that the cross-section 
is Thomson cross-section rather than exact Klein-Nishina cross-section.
With the energy enhancement factor $\eta$, the local radiative 
cooling rate $Q_{\rm rad}^{-}$ is given by
\begin{equation}
	Q_{\rm rad}^-=\int {d\nu \eta \left( \nu  \right)2F_\nu }.
	\label{eq2.36}
\end{equation}
We calculate the radiative cooling rate numerically and use it for the 
energy equation for electrons at each radius to calculate the global 
solution of the accretion flows.

\subsubsection{Calculation of the Spectrum}

We need more detailed specification for the calculation of the spectrum.  
Knowing the spectrum of unscattered photons and the probability that a 
photon will suffer scattering given times, we can calculate the Compton 
scattered spectrum if we know the information about how the original spectrum is 
modified after experiencing one scattering.  Note that the average energy 
boost given in (\ref{eq2.31}) is the energy boost {\it averaged} over the Maxwellian 
distribution of electrons and can not be used directly for the spectrum 
calculations.  Such problem was precisely discussed first by 
Jones (1968) and corrected 
afterwards by Coppi \& Blandford (1990).
We make use of the formula given by Coppi \& Blandford (1990) and calculate the 
resulting spectrum. 
The remaining problem concerning Compton 
scattering is how to treat the spectrum of the saturated Comptonized 
photons.  We assume that photons which are scattered more than $j_{m}$ times 
saturate and obey the Wien distribution 
$\propto \nu ^3\exp \left( {-h\nu / kT_e} \right)$.

Since large fraction of the emitted radiation is generated at the radius 
fairly close to the black holes in our model, we cannot ignore the effect 
of redshift due to the gravity and the gas motion. We include the gravitational 
redshift by simply taking the ratio of the energy of a photon when observed to 
its energy emitted at radius r to be $\sqrt {1-\rg/ r}$.
To treat redshift due to relativistic gas motion in a simple way, 
we concentrate on the face-on case where the optically thin assumption is most 
adequate.  Thus we simply take the ratio of energy change for the redshift due to 
gas motion to be $1/ \sqrt {1-(v/ c)^2}$.

\subsection{Numerical Procedure}

We solve numerically the set of equations given so far with the boundary 
conditions. The outer boundary conditions imposed are 
\begin{equation}
	\Omega =0.8\OmegaK,
	\label{eq2.36.5}
\end{equation}
\begin{equation}
	T_{\rm gas}\equiv
	\mu \left( {{{T_i} \over {\mu _i}}+{{T_e} \over {\mu _e}}} \right)=
	0.1T_{\rm vir},
	\label{eq2.36.75}
\end{equation}
\begin{equation}
	Q_{\rm rad}^-=\Lambda _{ie}
	\label{eq2.37}
\end{equation}
at $r_{\rm out}=10000\rg$. Here $T_{\rm vir}$ is the virial temperature
defined by
\begin{equation}
	T_{\rm vir}\equiv\left( {\gamma -1} \right){{G\MBH m_{\rm H}} \over {kr}}
	\label{eq2.38}
\end{equation}
We did not set the 
angular velocity to be the Keplerian angular velocity itself at the outer 
boundary for simply technical reasons.
We need to set somewhat sub-Keplerian disk at the outer boundary so 
that the viscous heating term take positive value with
the simple viscosity prescription adopted in this paper [eq. 
(\ref{eq2.9})].
We confirmed that the outer boundary condition have little effect on
the structure of the inner advection-dominated flows.   
There have been many suggestions about how outer Keplerian disks are 
connected to the advection dominated disks.  For instance, Honma  
(1996) took into account the effect of thermal conductivity and obtained 
the global solutions where the outer cool Keplerian disk is connected to 
the inner hot optically thin disk.  However the mechanism of the transition 
is not yet clear.  

The free parameters in the set of equations are $\alpha$, $\beta$, 
$\MBH$, $\dot M$, and $l_{\rm in}$.
These five parameters are not independent because of the 
transonic nature of the equations. $l_{\rm in}$ can be determined uniquely so that the 
solution should satisfy the transonic condition when the remaining parameters 
are given. We have to adjust $l_{\rm in}$ recursively to obtain smooth transonic 
solutions, since the location of the transonic point is not known 
until we obtain global transonic solutions.  

\section{Results}

To discuss the general properties of our model,
we set $m\equiv\MBH/\Ms=10^8$ and $\dot m\equiv\Mdot/\Mdotc=10^{-4}$
where $\Mdotc\equiv32\pi c\rg/ \kappa _{\rm es}$
and assign typical values for other parameters: 
$\alpha=0.1$, $\beta=0.5$ (case of equipartition).
To demonstrate how the heatings and coolings balance in the energy equations, 
we show Figure 1 the $Q$'s for the ions (upper panel), and for the electrons 
(lower panel); (see eq. [\ref{eq2.23}]).
The solid line in the upper panel shows the ratio of advective cooling of ions 
to the viscous heating, which is commonly denoted by $f$.  We have $f=1$ for the 
purely advection dominated accretion flows and $f=0$ for the purely radiative 
cooling dominated accretion flows.  We see that $f$ asymptotically approaches 
unity as the radius decreases.
We use the region $r<2000 \rg$ to investigate the 
spectrum from the optically thin advection dominated accretion flows.
We see from Figure 1 that the accretion flows becomes highly 
advection dominated in the region $r<100 \rg$ where the heat transport from ions to 
electrons by the Coulomb coupling practically becomes zero.  The interesting point 
is that it is electron advective {\it heating},
rather than the heat supplied from the ions, that balances with the 
radiative cooling of electrons, which means the electron accretion 
flow becomes cooling flow near the central black hole where
the electrons are cooled by consuming the stored internal energy
to radiate before being heated up by the ions (cf. Nakamura et al. 1997). 
Many analyses concerning two-temperature accretion flows have 
adopted simplified energy equation for electrons: $\Lambda 
_{ie}=Q_{\rm rad}^-$. 
Our results shows that above simplified energy equation is inadequate for the 
analysis of two-temperature advection dominated flows and one should adopt 
$Q_{{\rm adv},e}^-=-Q_{\rm rad}^-$ instead.

Figure 2 shows the ion and electron temperature profiles.  We see that the 
electron temperature is heated up to $T_{e}\sim10^{10}{\rm K}$.
Although we need to include the effect of electron-positron pair production 
and annihilation for such hot accretion flows, we make simple assumption 
that the pair density is very low (see discussion in Esin et al. 1996, 
Bjornsson et al. 1996, Kusunose \& Mineshige 1996).

Figure 3 shows the aspect ratio $h/r$ of the accretion flow.
We have used height-integrated equations for our calculations which neglect 
the higher order of $h/r$.  We may conclude that the value of $h/r$ shown in 
Figure 3 is marginally 
safe for the height integration.  However, of course, the vertical structure 
of the advection dominated accretion flows is an important issue and we 
will investigate it in future papers.

Figure 4 shows the angular momentum and the various velocities as a function of the 
radius $r$.  Here $c_{\rm s}^{*}$ is defined as 
\begin{equation}
	c_{\rm s}^{*}\equiv\left( {{{\left( {3\gamma -1} \right)+
	2\left( {\gamma -1} \right)\alpha ^2} \over {\gamma +1}}
	{W \over \Sigma }} \right)^{1/ 2},
	\label{eq3.1}
\end{equation}
so that $v_{r}=c_{\rm s}^{*}$ at the critical point.
We see from the Figure 4 that the radial dependence of the radial and the 
azimuthal velocities are different from that of the self-similar 
solutions, in which all the velocities are proportional to $r^{-1/2}$, 
especially in the super-sonic region. 
We see that for the parameters given above, 
the azimuthal velocity $v_{\varphi}$ is fairly 
sub-Keplerian and even sub-sonic and of the 
same order as the radial velocity $v_{r}$ unlike the standard model,
which is known to be a common feature for $\alpha=0.1$.
(In the standard model, we have $v_\varphi \sim r \OmegaK = c_{\rm 
s}^{*}h/ r\gg c_{\rm s}^{*}\gg v_{r}$.)
Note that we have `sub-sonic' azimuthal velocity when the rotation is
highly sub-Keplerian ($v_\varphi < r \OmegaK$) and the disk is hot and 
geometrically thick ($h/r\sim 1$).
For the reason that the radial velocity $v_{r}$ is the same order as the 
azimuthal velocity $v_{\varphi}$,
we call the accreting gas ``accretion flow'' rather 
than to call ``accretion disk'' in this paper.

Figure 5 illustrates the luminosity distribution as a function of $r$.  The 
contributions from respective radiation mechanisms (i.e.  bremsstrahlung, 
synchrotron, Comptonization) are also shown.  We see that the bremsstrahlung 
emission is important and the effect of the 
synchrtron emmision and the 
Comptonization is negligibly small in the outer region of the accretion 
flow ($r>10\rg$), while the synchrotron emission and the Comptonization dominate in the 
hot inner region ($r<10\rg$). 
The contribution of the Comptonization rapidly increases as the 
radius decreases because the amount of the synchrotron soft photons 
increases and not because the Compton $y$-parameter $y\equiv\tau_{\rm 
es}kT_{e}/m_{e}c^{2}$ increases. Note that in the innermost region 
the electron temperature $T_{e}$ is approximately constant and the 
surface density decreases and thereby $y$ actually decreases.
Note that almost all emission comes from the hot inner 
region ($r<10\rg$) and the fraction of the emission from the super-sonic 
region, which the self-similar solutions fail to describe accurately,
is fairly large. As far as the 
bremsstrahlung is concerned, the contribution of the emission from the 
large radii is not negligible, which make the slope of the 
bremsstrahlung spectrum less steep, since what we observe is the 
superposition of the bremsstrahlung peaks from different radii.

In Figure 6, we shows the spectrum generated by the optically thin 
accretion flows.  
Parameters are $m=10^8$, $\dot m=10^{-4}$,
$\alpha=0.1$, $\beta=0.5$.
S indicates the synchrotron 
peak which is composed of Rayleigh-Jeans slope and the optically thin 
synchrotron emission.  C1 and C2 indicate the once and twice Compton 
scattered photons, respectively.  B indicates the bremsstrahlung 
emission plus 
photons suffering multiple Compton scattering.  W indicates saturated 
Comptonized photons which form Wien tail. 

The upper panel of Figure 7 shows the surface density of the accretion flow.  The dashed line 
corresponds to the case with the central black 
hole mass of $\MBH=10\Ms$.  We find that the structure of the accretion 
flow is nearly the same when 
the radius and the mass accretion rate are 
scaled with the Schwarzschild radius and the critical 
mass accretion rate, respectively.  However we see in the lower panel 
of the Figure 7 that the electron temperature at the innermost region 
is varied slightly.  
This is due to the fact that the accretion flow is highly 
advection dominated.
We remind the readers that the viscous heating is balanced almost entirely 
with the advective cooling in the ion energy equation and there is little 
coupling between the ions and the electrons.  Thus the structure of the 
accretion flow is governed by ion energy equation while the electron energy 
equation including the radiative cooling is decoupled and determines the 
electron temperature.  Although the height-integrated quantities are the 
same when the radius is scaled with the Schwarzschild radius, the amount of radiative 
cooling, which is the function of the density distribution rather 
than the height-integrated quantities, is 
different.  Thus we have different electron temperatures for different black 
hole masses.
We show the spectrum from the accretion flow for the $10\Ms$ case in 
Figure 8. The shape of the spectrum is essentially the same as the 
$10^{8}\Ms$ case, but the position of the synchrotron peak and the 
absolute luminosity differs considerably.

\section{Application to Sagittarius A*}

Following up on the successful work of Narayan et al. (1995)   which applied the 
advection dominated model to the Sgr A* (the central core of our Galaxy), 
we improve their model 
by fully solving the basic equations to explain the observed emission from radio 
frequencies to Gamma-rays.  
Figure 9 shows the model which explains observed radio and X-ray data 
quite well. 
The parameters assigned are given in the 
figure.  The points and the short lines are the observational data, 
which assume interstellar absorption ${\rm N}_{H}=6\times10^{22}cm^{-2}$
and a distance $d=8.5{\rm kpc}$, both typical for the Galactic Center,
compiled by Narayan et al. (1995) (references therein).  
The various lines correspond to the 
different values of mass accretion rate which varies by factor of 2.
As Narayan et al.  noticed, the position of the Rayleigh-Jeans slope is 
determined solely by the mass of the central black hole.  
Thus to give a good fit to the 
observed radio emission with our model, we have no choice but to fix the mass of the 
central black hole to be $\MBH=10^{6}\Ms$. 
Considering the inaccuracy of the potential at 
the innermost region of the accretion flow, this value is consistent with 
the value which is derived from the gas and stellar dynamics 
(Genzel \& Townes 1987). We have $\Mdotc=3.5\times10^{-2}\Ms/{\rm yr}$
for the black hole of the mass $\MBH=10^{6}\Ms$,
hence the predicted mass accretion rate is $\sim 2-5 \times 10^{-6}\Ms/{\rm yr}$.

Figure 9 also illustrates the $\Mdot$ dependence of our model.  The surface 
density sensitively depends on $\Mdot$, while the temperature is fairly 
insensitive to $\Mdot$.  Thus the luminosity at all frequencies decreases 
when we reduce the mass accretion rate.
We find that if we change the mass 
accretion rate by a factor of $\sim2$, X-ray luminosity varies by the same factor 
while the synchrotron peak varies little.  We suggest that the various 
X-ray data seemingly inconsistent with each other are purely due to the 
change of the mass accretion rate of the accretion flow.
There have been an argument that the X-ray luminosity of Sgr A* is variable 
on the timescale of half a year, which is consistent with our suggestion.

Figure 10 illustrates 
how the spectrum changes with different black hole masses.  As we mentioned 
before, the structure of the accretion flow is the same with proper 
scalings, but the electron temperature is slightly different because of the 
difference of the emissivity.
We see clearly that the position of the Rayleigh-Jeans slope is 
determined by the mass of the central black hole.

We show in Figure 11 the $\beta$-dependence of the structure of the accretion 
flow and the spectrum.  We remind the readers that the parameter $\beta$ is 
defined as the ratio of the gas pressure to the total pressure.  We have 
the stronger magnetic field for the smaller value of $\beta$.
Unlike the other parameter dependence, the ion temperature as well as the 
electron temperature decreases when we lower the value of $\beta$.  This is 
reasonable because for the lower value of $\beta$, the stronger becomes the magnetic 
pressure and the gas pressure can be smaller to support the accretion flow.
This temperature change has a large effect on the bremsstrahlung spectrum but 
has little effect on the synchrotron peak, which is because the effects of 
the stronger (weaker) magnetic field and the lower (higher) temperature 
roughly cancel out.
We set $T_{\rm gas}=0.2T_{\rm vir}$ at the outer boundary for 
$\beta=0.95$ since we 
could not find advection dominated solution for $\beta=0.95$ with 
$T_{\rm gas}=0.1T_{\rm vir}$.

Narayan et al. (1995) did not investigate the parameter dependence of the viscosity 
parameter $\alpha$ and the mass accretion rate $\Mdot$ independently, since 
they used simplified disk model, in which $\alpha$ and $\Mdot$ 
cannot be chosen independently.  
Thus it is meaningful to study the $\alpha$ dependence of the spectrum.  We 
show in Figure 12 the $\alpha$-dependence of our model.
When $\alpha$ is large, the angular momentum is 
extracted efficiently and the surface density decreases, which makes the 
bremsstrahlung emission weak.
However, the increase of the electron temperature at the innermost region 
makes the synchrotron emission stronger.  We find that the width of the 
synchrotron peak is the most sensitive to the value of $\alpha$.  The upper 
limits in the radio frequency band imposes strong restriction upon the 
value of $\alpha$ for the case of Sgr A*.  For instance, we can not fit the 
entire spectrum with $\alpha=0.1$.  We have to set $\alpha<0.025$ to explain 
entire spectrum of Sgr A* with our model. This fact is important 
because it is claimed that $\alpha\sim0.1$ in advection-dominated 
accretion flows on various grounds (see discussions in Narayan 1996).
If we are to set $\alpha=0.1$, we have to reduce the mass accretion 
rate by factor of 4 so as to give a good fit to the observed radio and 
IR spectrum (see Figure 12). 
In that case, we cannot explain the X-ray emission from 
the Galactic Center. Various X-ray observations have found the X-ray 
sources at the Galactic Center but the angular resolution has been 
insufficient to identify a source with the radio source Sgr A*.
There remains the possibility that X-rays do not come from Sgr A* at all 
(e.g. see Duschl et al. 1996). If that is the case, all the X-ray 
data are merely upper limits and the accretion flows with $\alpha=0.1$ 
do not conflict with the observations.

There exist some data points in near IR and radio frequencies
which cannot be accounted for.  However, it is natural to think 
that there must be dust region or stellar contamination or non-thermal 
objects like jets in central region of our Galaxy overlapping the 
accreting black hole which in total we observe as a 
point source Sgr A*.  In that sense, one should consider the observational 
data as upper limits.
A point worth emphasizing here is that the entire spectrum of Sgr A* is 
basically explained with an optically thin accretion flow around a black 
hole of the mass $\MBH=10^{6}\Ms$.  Moreover, we have solved globally the basic 
equations including the gas dynamics and the radiation processes 
consistently rather than to consider more primitive models like isothermal 
gas complex of certain size.

\section{Summary and Discussion}

In this paper, we have calculated the global structure of optically 
thin advection dominated accretion flows in the context of two-temperature 
plasma, adopting the full set of basic equations including the energy 
equation for the electrons. We have also calculated the spectra 
emitted by the optically thin accretion flows which we calculated.
We have made the calculation of the spectra and that of the structure of the 
accretion flows to be completely consistent by calculating the 
radiative cooling rate at each radius by numerically integrating 
the whole spectrum emitted at the radius.

As a result of the advection domination for the ions, the heat 
transport from the ions to the electrons becomes practically zero and 
the radiative cooling balances with the advective {\it heating} of 
the electrons. This means that the electron cools itself by releasing 
the stored internal energy as a radiation. Hence the energy equation
for the electrons play an important role for the calculation of the 
spectra, where the temperature profile of the electron is the 
important factor.

An interesting feature of the advection dominated flow,
which is known already, is that the 
azimuthal velocity becomes 
highly sub-Keplerian and of the same order as 
the radial velocity and the sound velocity.
The point worth noting is that in the innermost hot luminous region, 
the divergence of the velocities from those in the self-similar solution 
is fairly large.

The accreting gas becomes very hot. The electron temperature even exceeds 
the rest mass energy of an electron. We have not taken into account 
the effect of the pair production and the annihilation, which is 
an important issue. For such hot accretion flows, the synchrotron 
emission and the Compton scattering are very important.

The spectrum is composed by 1) the synchrotron peak which comprises 
optically thin synchrotron emission and the self-absorbed 
Rayleigh-Jeans slope and 2) the unsaturated Comptonized photons which 
forms some bumps and 3) the bremsstrahlung emission and 4)the 
saturated Comptonized photons. The dependence of the each component on 
the model parameters is complex. Among them, the position of the 
Rayleigh-Jeans slope is almost solely determined by the mass of the 
central black hole. When we make the magnetic field stronger, the 
temperature of the entire flow decreases, which has significant effect 
on the Comptonization and the bremsstrahlung emission, but has little 
effect on the synchrotron emission. When we make the viscosity 
smaller, the surface density increases and the bremsstrahlung emission 
increases, but the synchrotron emission and the Comptonization 
decreases. The simplest relation is the dependence on the mass 
accretion rate. When we reduce the mass accretion rate, the entire 
emission is reduced. However the bremsstrahlung emission is much more 
sensitive to the change of the mass accretion rate than the 
synchrotron emission.

We find that the spectrum of Sgr A* is explained by the optically thin 
advection dominated accretion flow around a black hole of the mass 
$\MBH=10^{6}\Ms$. Narayan et al. (1995) also calculated the spectrum of 
Sgr A* using an optically thin advection dominated accretion flow model.
Their best fit parameters are different from ours. 
For instance the mass of the central black hole is $\MBH=7.0\times 
10^{5}\Ms$ according to their model.
The different points in our model are 1) full global treatment of the 
basic equations and 2) inclusion of the electron energy equation and 
3) calculation of the innermost region where the flow is supersonic 
and the effects of the redshift are important. 2) and 3) have very 
important effect on the emitted spectrum, which is not considered in 
Narayan et al. (1995).
We conclude that the X-ray data obtained by various satellite 
observations are explained by the variation of the mass accretion 
rate by a factor of $\sim 2$, if we allow $\alpha$ to have small value.
If we set $\alpha\sim0.1$, which is considered to be a standard value
for the advection dominated accretion flows, it is not likely that 
the X-rays come from Sgr A*.

We have computed the model using height-integrated equations with fixed 
structure in the vertical direction. We have also adopted simplified 
form of equations for the radiation field. Our immediate goal is to 
solve the basic equations including equations for the radiation field
in at least two dimensional space.
However, our successful result presented in this paper tells us that the 
basic idea is correct.
The full treatment  of Schwarzschild or Kerr metric is also an 
important issue.

\acknowledgments

We thank Professor Shoji Kato for useful discussions, 
and Ann Esin and Jun Fukue for useful comments. We are very grateful to 
R. Narayan for providing us with a code to calculate
inverse-Compton spectra. We thank the referee J. P. Lasota for 
many valuable comments and suggestions.

\appendix

\section{Height Integrations of the Basic Equations}

In this paper, we have adopted the height-integrated equations to 
obtain the structure of the optically thin advection dominated accretion 
flows. We have seen that the temperature of the accretion flows is 
very high and the aspect ratio $h/r$ is $\sim 0.5$. 
Thus the height-integration may not be an excellent approximation since
the operation of 
the height-integration omits the higher orders of $h/r$.
Thus we consider it useful to describe in detail the operation of the 
height-integration to discuss the reliability of our calculations.

The potential we adopted is spherically symmetric, while we adopt the 
cylindrical coordinate system. Thus we need to expand it around the 
equatorial plane ($z=0$) in order to integrate the basic equations
in the vertical direction.
In the case of geometrically thin accretion flows, we can neglect 
the terms containing the higher orders of $z/r$ and the operation of 
the height-integration becomes very simple.

The potential $\psi$ is expanded as
\begin{equation}
	\psi \left( {r,z} \right)=\psi \left( {r,0} \right)\left[ {1+{1 \over 
	2}\left.  {{{\partial \ln \psi } \over {\partial \ln r}}} 
	\right|_{z=0}{{z^2} \over {r^2}}+O\left( {{{z^4} \over {r^4}}} \right)} 
	\right].
	\label{eqA.1}
\end{equation}
Thus we approximate $\psi$ as
\begin{equation}
	\psi \left( {r,z} \right)=\psi \left( {r,0} \right)+{1 \over 2}
	\Omega _k^2z^2,
	\label{eqA.2}
\end{equation}
where $\Omega_{k}$ is defined as
\begin{equation}
	\Omega _k\equiv \left.  {\left( {{1 \over r}{{\partial \psi } \over 
	{\partial r}}} \right)^{1/ 2}} \right|_{z=0}.
	\label{eqA.3}
\end{equation}
Using Equation (A.1), the derivatives of $\psi$ are also expanded as 
follows:
\begin{equation}
	{{\partial \psi } \over {\partial r}}=\left.  {{{\partial \psi } \over 
	{\partial r}}} \right|_{z=0}\left[ {1+{{d\ln \Omega _k} \over {d\ln 
	r}}{{z^2} \over {r^2}}+O\left( {{{z^4} \over {r^4}}} \right)} \right],
	\label{eqA.4}
\end{equation}
\begin{equation}
	{{\partial \psi } \over {\partial z}}=\left.  {{{\partial \psi } \over 
	{\partial r}}} \right|_{z=0}{z \over r}\left[ {1+O\left( {{{z^3} \over 
	{r^3}}} \right)} \right].
	\label{eqA.5}
\end{equation}
We approximate the derivatives of $\psi$ as follows:
\begin{equation}
	{{\partial \psi } \over {\partial r}}=r\Omega _k^2\left( {1+{{d\ln \Omega 
	_k} \over {d\ln r}}{{z^2} \over {r^2}}} \right),
	\label{eqA.6}
\end{equation}
\begin{equation}
	{{\partial \psi } \over {\partial z}}=\Omega _k^2z.
	\label{eqA.7}
\end{equation}
We have used eq. (\ref{eqA.7}) to determine the vertical structure of the 
accretion flows [see eq. (\ref{eq2.15})].

We then integrate the basic equations in the vertical direction. The 
integrations of the continuity equation (eq. [\ref{eq2.5}]) and the azimuthal 
component of the Euler equation (eq. [\ref{eq2.8}]) are straightforward. In 
the process of the integration of the radial component of the Euler 
equation, we encounter the following integration:
\begin{equation}
	\int_{-\infty }^\infty {\rho {{\partial \psi } \over {\partial 
	r}}}dz=\Sigma r\Omega _k^2+W{{d\ln \Omega _k} \over {dr}}.
	\label{eqA.8}
\end{equation}
The last term in the right hand side of eq. (\ref{eqA.8}) is sometimes
omitted in the height-integrated equations which appear in the papers 
on the accretion disks, which is not proper, since it can be
the same order as the other terms in the Euler equation.

The left hand sides of the energy equations (eq. [\ref{eq2.10}], eq. [\ref{eq2.11}]) together with the
thermodynamic relations (eq. [\ref{eq2.13}], eq. [\ref{eq2.14}])
are transformed to give
\begin{equation}
	\rho T\left( {v_r{{\partial s} \over {\partial r}}+v_z{{\partial s} \over 
	{\partial z}}} \right)={\gamma \over {\gamma -1}}\left[ {{1 \over 
	r}{\partial \over {\partial r}}\left( {rv_rp} \right)+{\partial \over 
	{\partial z}}\left( {v_zp} \right)} \right]-v_r{{\partial p} \over 
	{\partial r}}-v_z{{\partial p} \over {\partial z}}.
	\label{eqA.9}
\end{equation}
We can integrate Equation (\ref{eqA.9}) with the integration
\begin{equation}
	\int_{-\infty }^\infty {v_z{{\partial p} \over {\partial 
	z}}}dz=-Wv_r{{d\ln H} \over {dr}},
	\label{eqA.10}
\end{equation}
to obtain the height-integrated version of the Energy equations (eq. 
[\ref{eq2.21}], [\ref{eq2.22}]).

\section{Calculation of The Flux of Unscattered Photons}

Consider the isothermal plane parallel atmosphere with the density 
distribution being
\begin{equation}
	\rho \left( z \right)=\rho \left( 0 \right)\exp \left( {-{{z^2} \over 
	{2H^2}}} \right).
	\label{eqB.1}
\end{equation}
Assuming the Eddington Approximation which is valid for isotropic 
radiation fields (and even for slightly nonisotropic fields, see 
Rybicki \& Lightman 1979), the radiation field in the vertical 
direction is described by the radiative diffusion equation:
\begin{equation}
	{1 \over 3}{{\partial ^2J_\nu } \over {\partial \tau _\nu ^2}}=J_\nu 
	-B_\nu ,
	\label{eqB.2}
\end{equation}
where $\tau _\nu$ is the optical depth from the surface of the 
accretion flow. There is no well-defined surface of the accretion 
flow since the density tends to zero 
when we increase $z$ but will never equals to zero [see eq. 
(\ref{eqB.1})]. However, the 
optical depth $\tau_{\nu}$ of the accretion flow in the vertical 
direction is finite and hence we can define the surface when the vertical 
height is measured with the optical depth. 
Note that $\tau=\tau _\nu ^*={{\sqrt \pi } \over 2}\kappa _\nu (0)H$
at the equatorial plane and $\tau=2\tau _\nu ^*$ at the other surface.
We solve eq. (\ref{eqB.2}) with boundary conditions. We take 
\begin{equation}
	\begin{array}{l}
	{1 \over {\sqrt 3}}{{\partial J_\nu } \over {\partial \tau _\nu 
	}}=J_\nu \quad(\tau _\nu =0),\\
	{{\partial J_\nu } \over {\partial \tau 
	_\nu }}=0\quad(\tau _\nu =\tau _\nu ^*).
	\end{array}
	\label{eqB.3}
\end{equation}
The boundary condition at the surface assumes there is no irradiation 
onto the surface of the accretion flow and is derived by adopting 
two-stream approximation (see Rybicki \& Lightman 1979).
The solution for eq. (\ref{eqB.2}) which satisfies the boundary conditions 
[eq. (\ref{eqB.3})] is
\begin{equation}
	J_\nu =B_\nu \left\{ {1-{{e^{-\sqrt 3\tau _\nu }} \over 2}\left[ 
	{e^{-2\sqrt 3\left( {\tau _\nu ^*-\tau _\nu } \right)}+1} \right]} 
	\right\},
	\label{eqB.4}
\end{equation}
The energy flux $F_{\nu}$ on the surface of the accretion flow is 
given by
\begin{equation}
	F_\nu \left( 0 \right)={{4\pi } \over 3}\left.  {{{\partial J_\nu } \over 
	{\partial \tau _\nu }}} \right|_{\tau _\nu =0}={{2\pi } \over {\sqrt 
	3}}B_\nu \left[ {1-\exp \left( {-2\sqrt 3\tau _\nu ^*} \right)} \right].
	\label{eqB.5}
\end{equation}
Note that
\begin{equation}
	\begin{array}{l}
	F_\nu \left( 0 \right)={{2\pi } \over {\sqrt 3}}B_\nu 
	\quad(\tau _\nu ^*\gg1),\\
	F_\nu \left( 0 \right)={{\sqrt \pi } \over 
	2}\chi _\nu \left( {z=0} \right)H\quad(\tau _\nu ^*\ll1).
	\end{array}
	\label{eqB.6}
\end{equation}

\clearpage

\figcaption{The amount of the heating and the cooling rates for the ions 
(upper panel) and for the electrons (lower panel) in the units of
$2 \pi r^{2}/(\Mdot W_{i}/\Sigma)$ and $2 \pi r^{2}/(\Mdot W_{e}/\Sigma)$,
respectively.
The ratio of advective cooling 
of ions to the viscous heating $f$ is also shown in the upper panel. The 
accretion flow becomes advection dominated ($f=1$) as the radius 
decreases and the energy transport from the ions to the electrons 
becomes practically zero. The radiative cooling is balanced with the 
electron advective {\it heating} in the innermost region. \label{fig1}}

\figcaption{The temperature of the ions and the electrons as a 
function of $r$. The electrons attain the temperature of $10^{10}{\rm 
K}$. \label{fig2}}

\figcaption{The aspect ratio $h/r$ of the accretion flow. \label{fig3}}

\figcaption{The angular momentum (upper panel) and the various 
velocities (lower panel) of the accretion flow. The azimuthal 
velocity $v_{\varphi}$ is highly sub-Keplerian. In the hot luminous 
region near the central black hole, the divergence of the velocities from 
those in the 
self-similar solution is fairly large. \label{fig4}}

\figcaption{The luminosity distribution as a function of $r$.  The 
contributions from respective radiation mechanisms (i.e.  bremsstrahlung, 
synchrotron, Comptonization) are also shown. The synchrotron emission 
and the Compton scattering dominate in the innermost region. The 
distribution of the bremsstrahlung emission is relatively flat. 
\label{fig5}}

\figcaption{The spectrum generated by the optically thin 
accretion flow around the central black hole of the mass 
$\MBH=10^{8}\Ms$. S indicates the synchrotron 
peak which comprises Rayleigh-Jeans slope and the optically thin 
synchrotron emission.  C1 and C2 indicate the once and twice Compton 
scattered photons, respectively and B indicates the bremsstrahlung 
emission plus 
photons suffering multiple Compton scattering.  W indicates saturated 
Comptonized photons.  \label{fig6}}

\figcaption{The surface density (upper panel) and 
the temperature (lower panel) of the accretion flow. The dashed line 
corresponds to the case where $\MBH=10\Ms$. We find that the structure of the accretion 
flow is nearly the same when 
the radius and the mass accretion rate are 
scaled with the Schwarzschild radius and the critical 
mass accretion rate, respectively, while the electron temperature at the innermost region is varied 
slightly.  \label{fig7}}

\figcaption{The spectrum generated by the optically thin 
accretion flow around the central black hole of the mass 
$\MBH=10\Ms$. The letters indicate the same meaning as Fig. 6.
The shape of the spectrum is essentially the same. However, the 
position of the synchrotron peak and the total luminosity differ 
significantly. \label{fig8}}

\figcaption{The spectrum of Sgr A* (upper panel). The lines correspond to the 
spectrum calculated with our model presented in this paper.
The $\Mdot$-dependence of the spectrum and the 
temperature and the surface density (lower panel) are illustrated.
The spectra behaves in the simplest way. When $\Mdot$ is reduced,  
the surface density and the entire emission are also reduced.
The bremsstrahlung emission is more sensitive to the  $\Mdot$ change 
than the synchrotron emission. 
\label{fig9}}

\figcaption{The $\MBH$-dependence of the spectrum (upper panel) the 
temperature and the surface density (lower panel).
The position of the Rayleigh-Jeans slope determines the mass of the 
central black hole. The surface density is not varied when $\MBH$ is 
changed. \label{fig10}}

\figcaption{The $\beta$-dependence of the spectrum (upper panel) the 
temperature and the surface density (lower panel).
When $\beta$ is lowered, the temperature decreases and the 
bremsstrahlung emission is weakened. However, the synchrotron emission 
is fairly insensitive to the change of $\beta$. \label{fig11}}

\figcaption{The $\alpha$-dependence of the spectrum (upper panel) the 
temperature and the surface density (lower panel).
When $\alpha$ is lowered, the surface density increases and so the 
bremsstrahlung emission. However, the synchrotron emission weakened 
due to the decrease of the electron temperature in the innermost 
region. \label{fig12}}

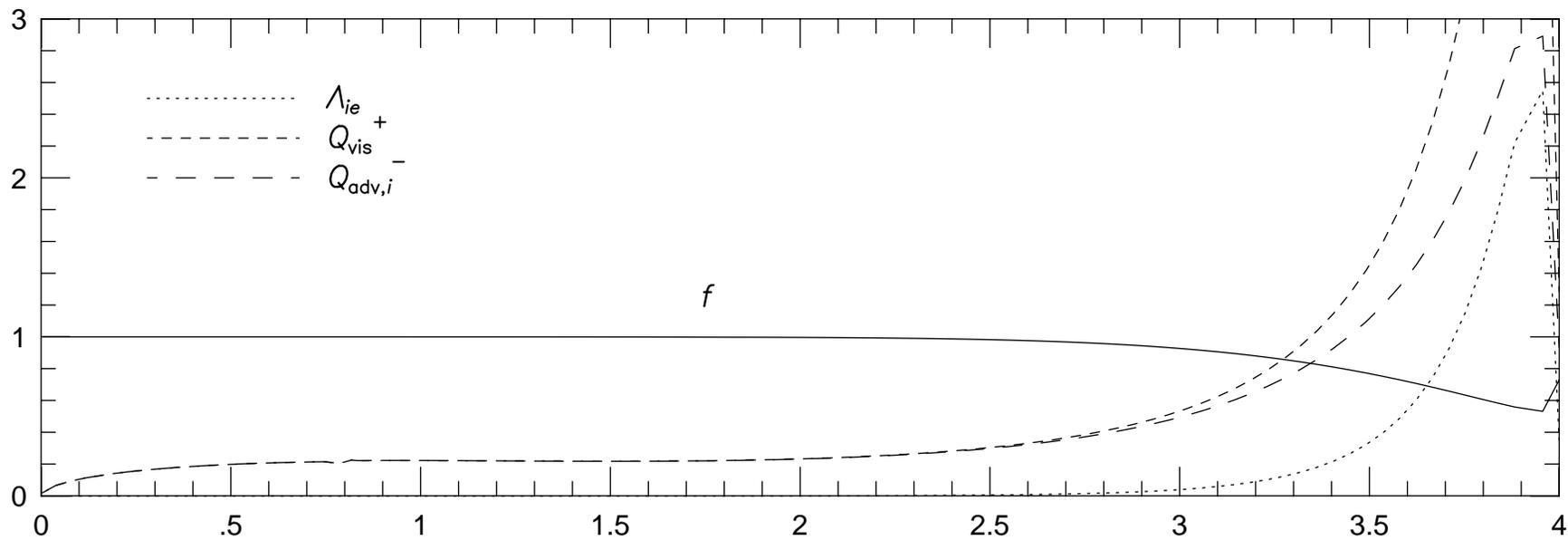
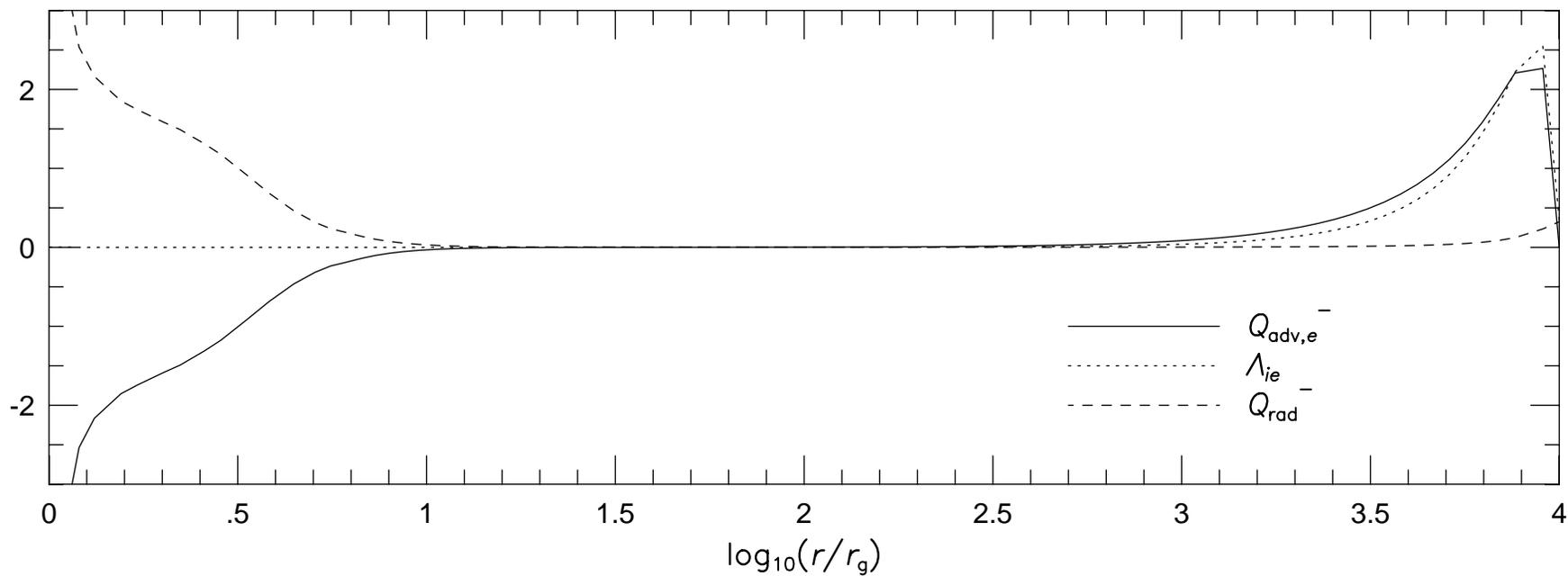
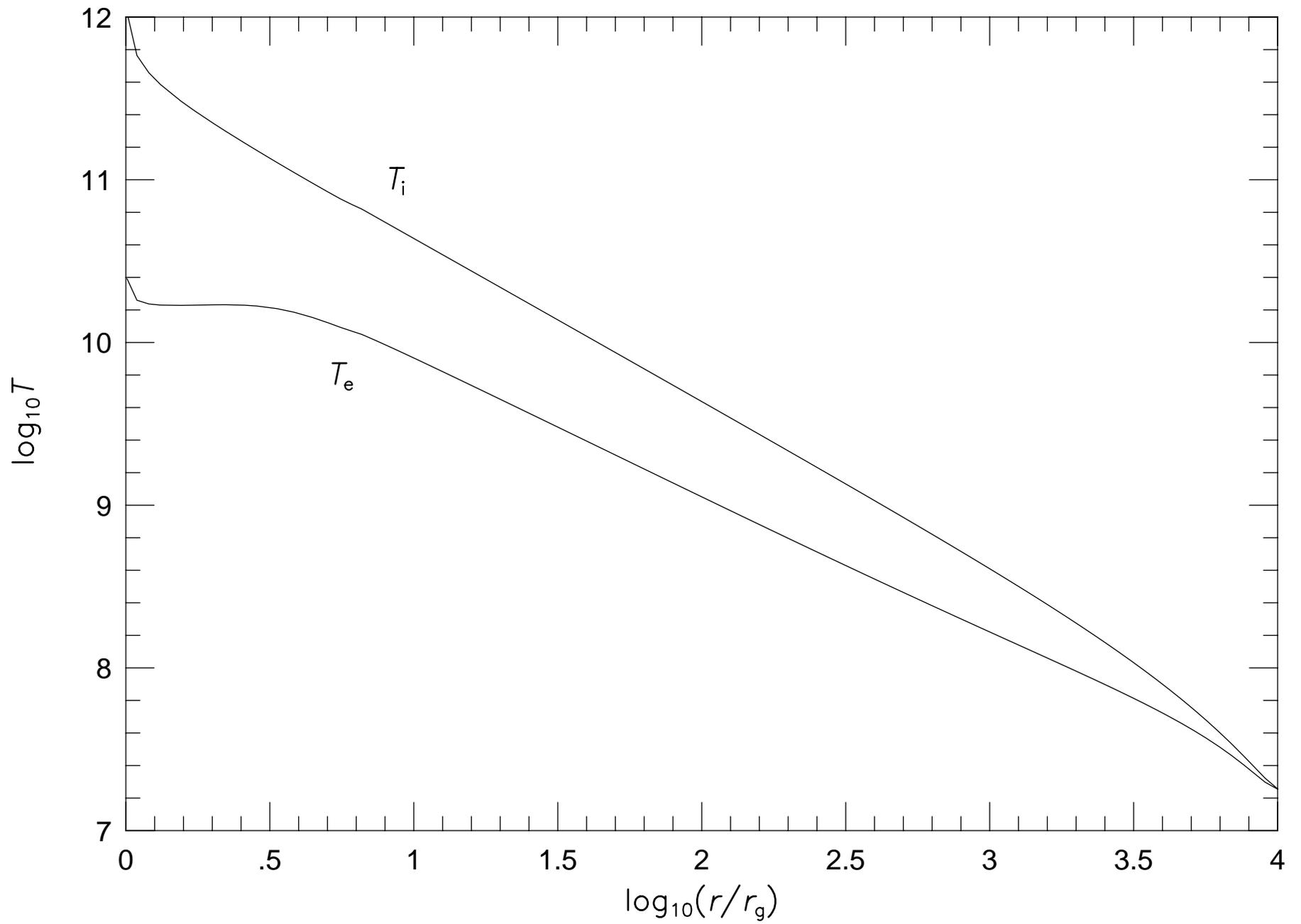
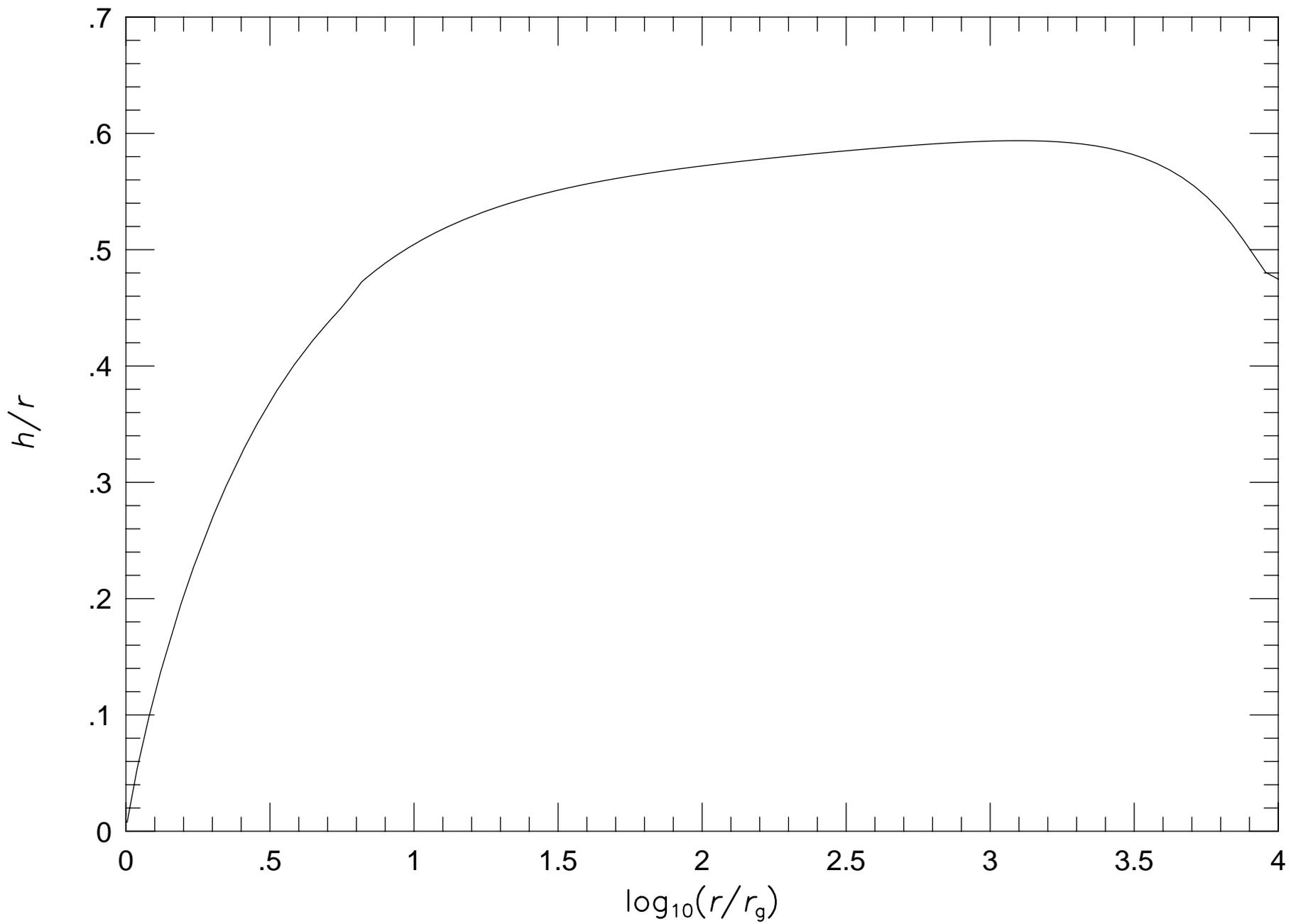
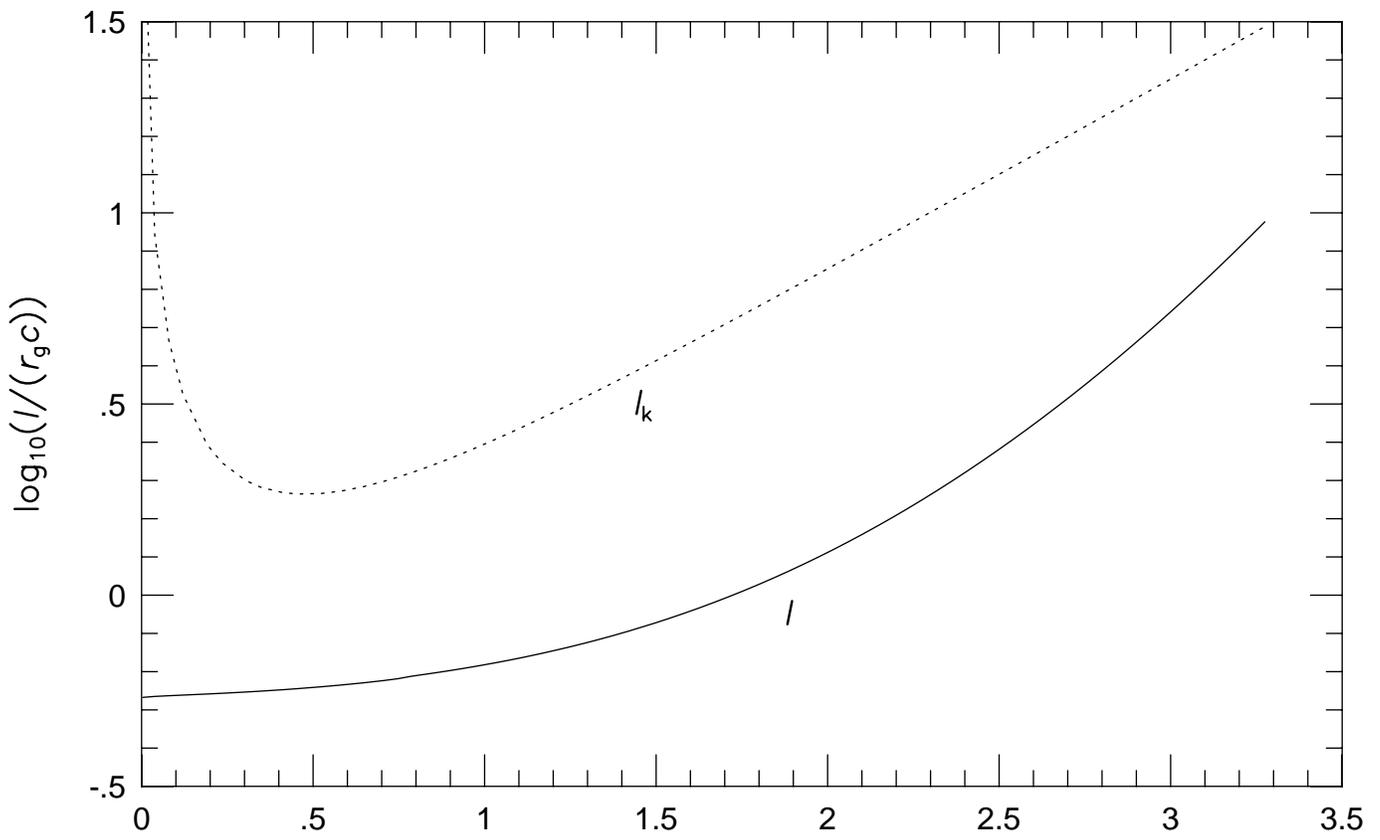
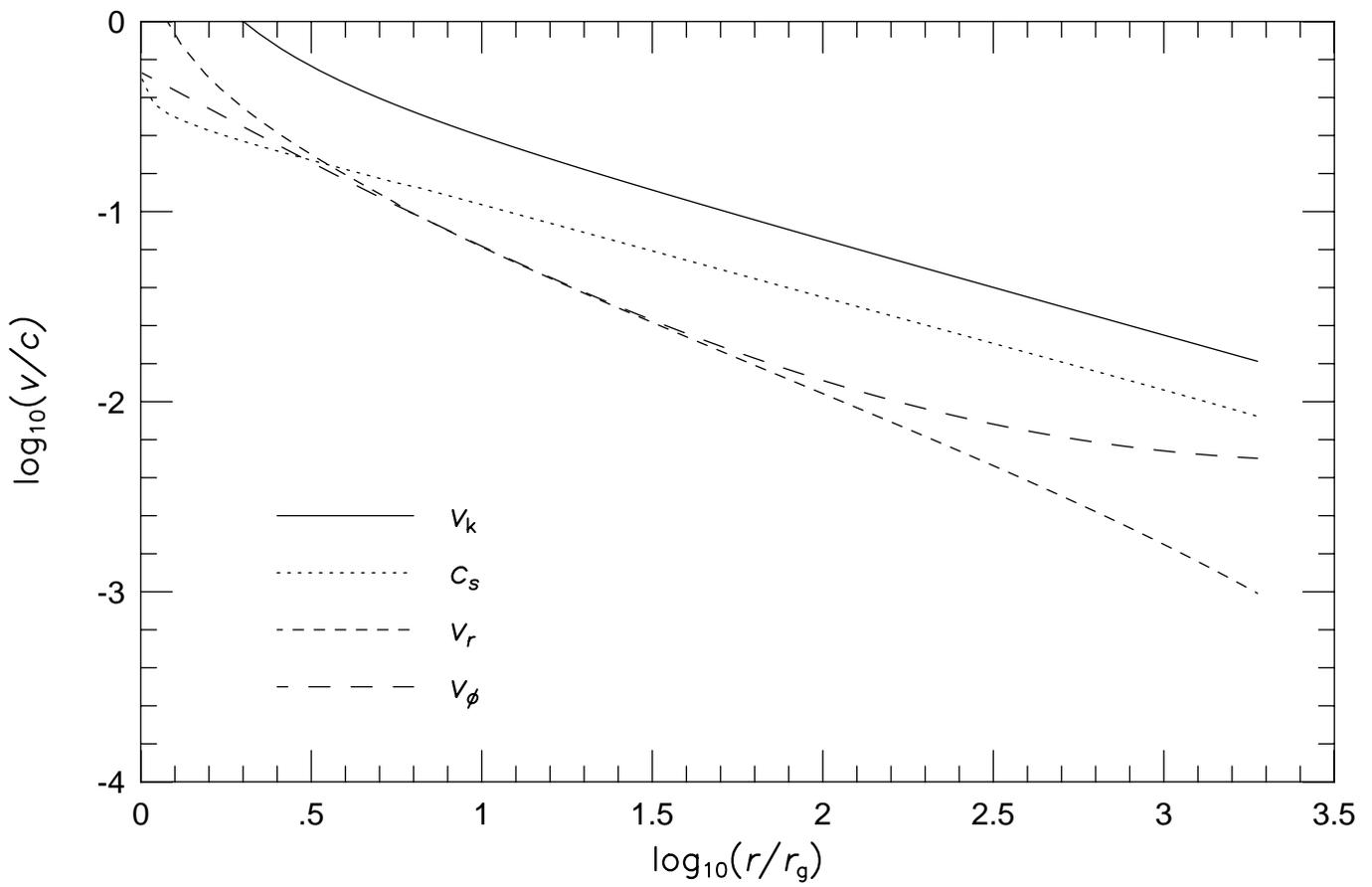
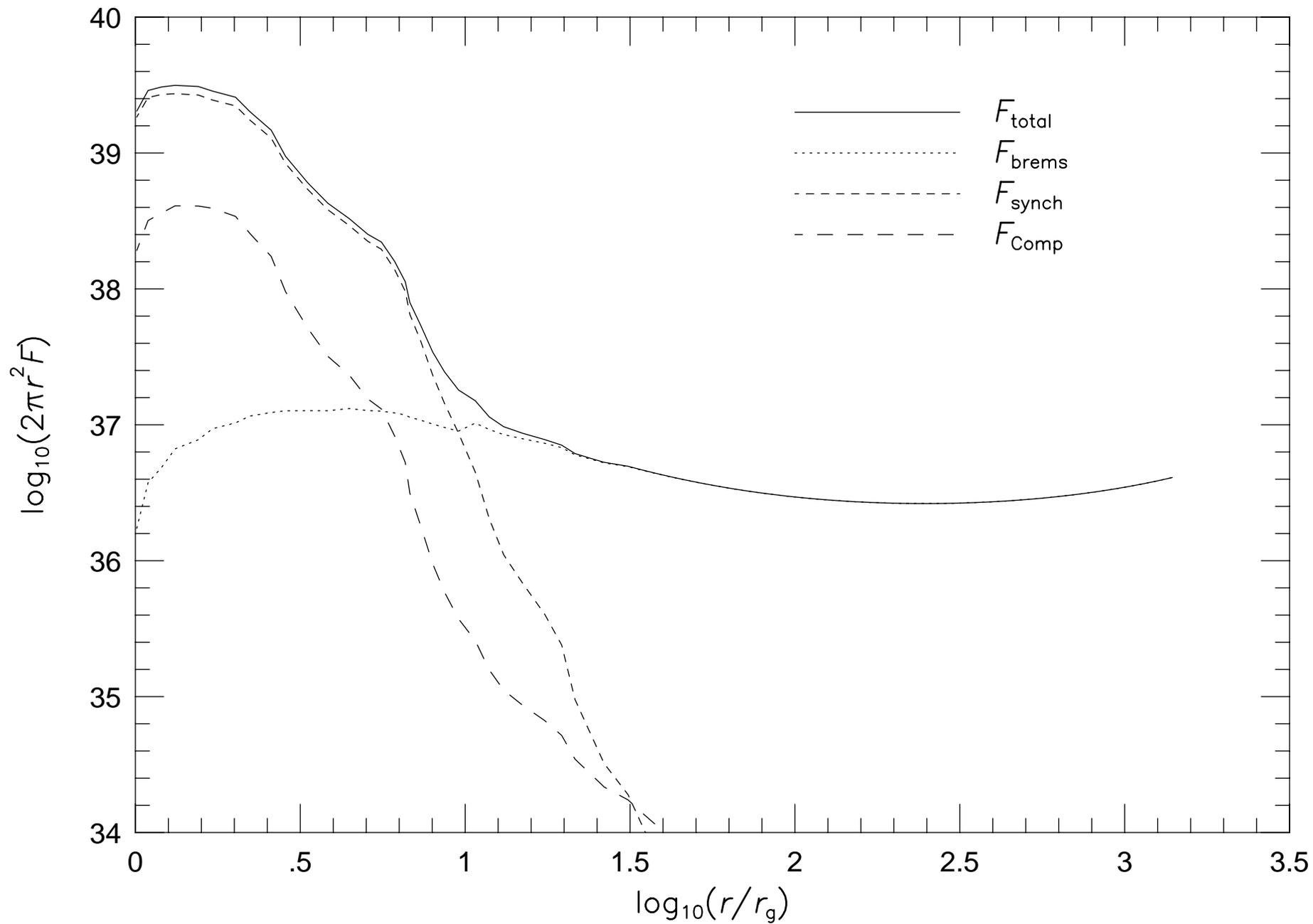
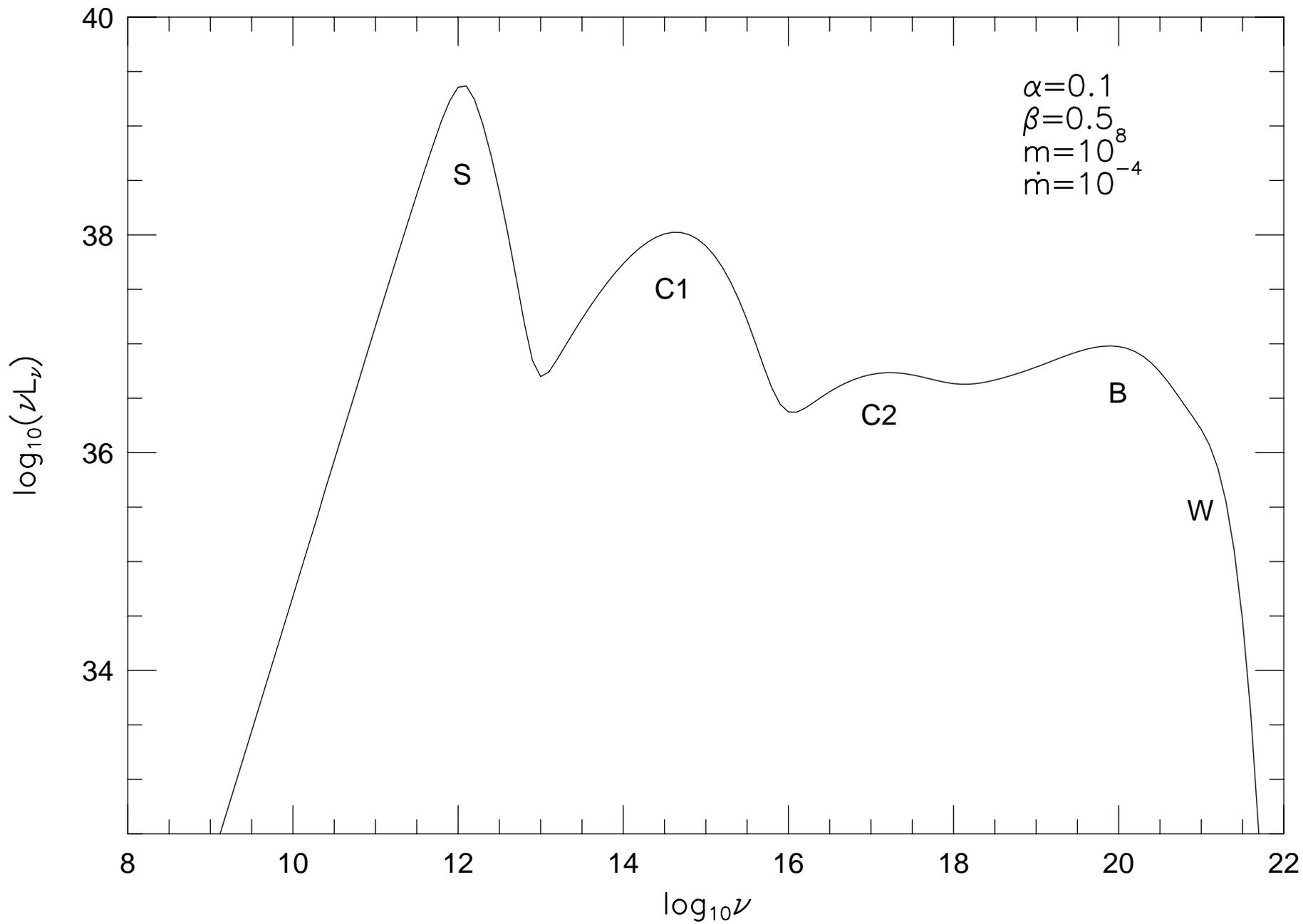
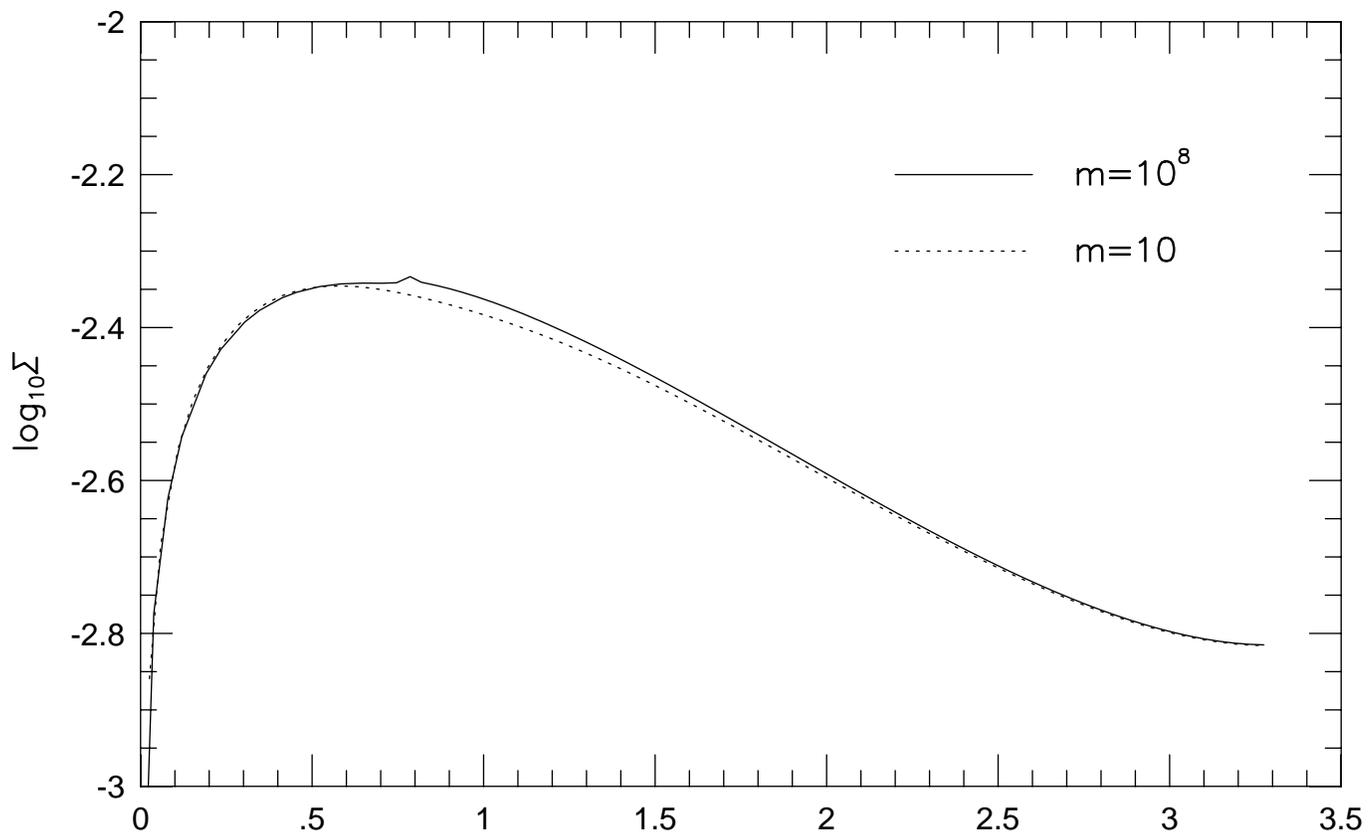
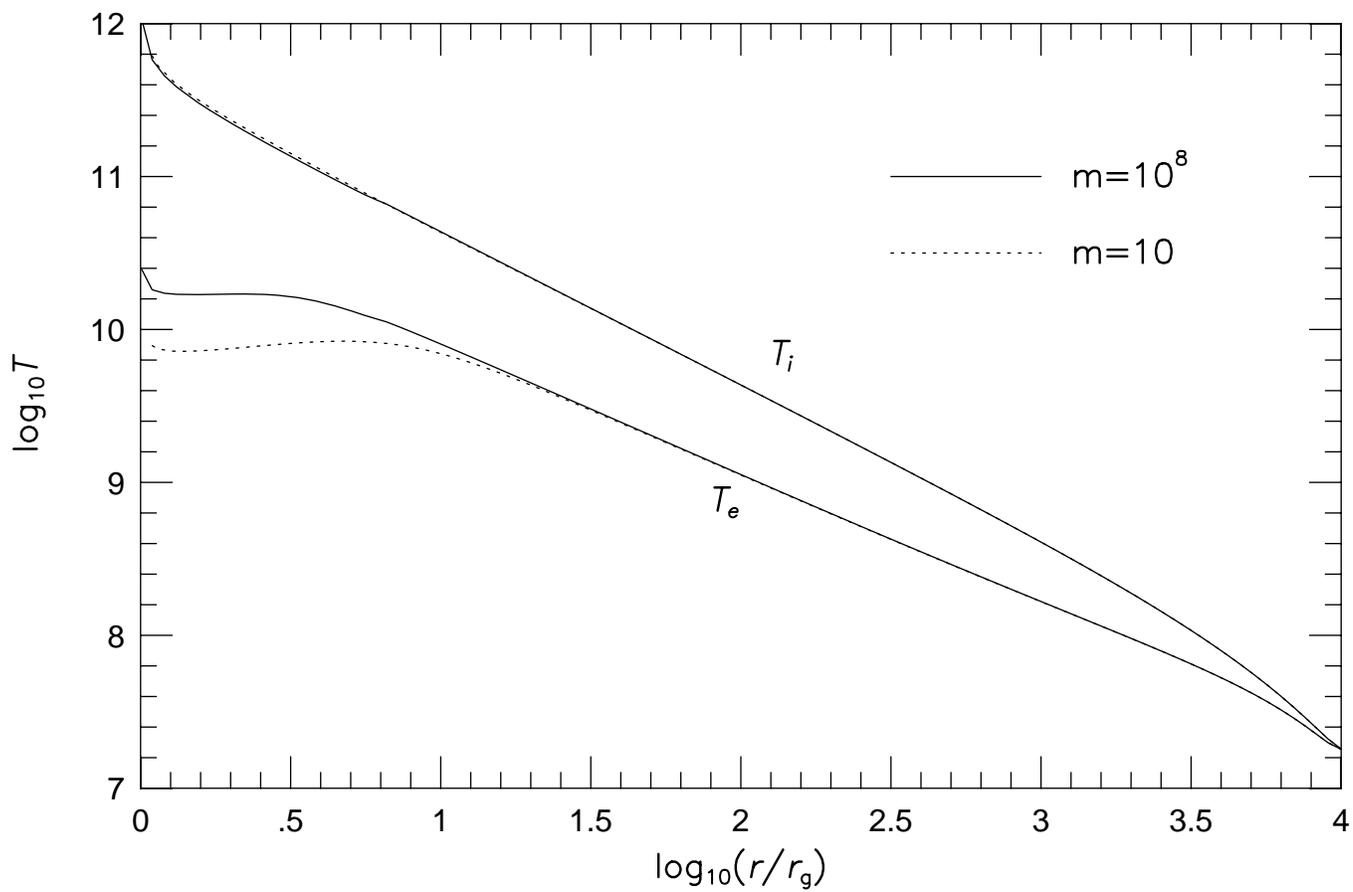
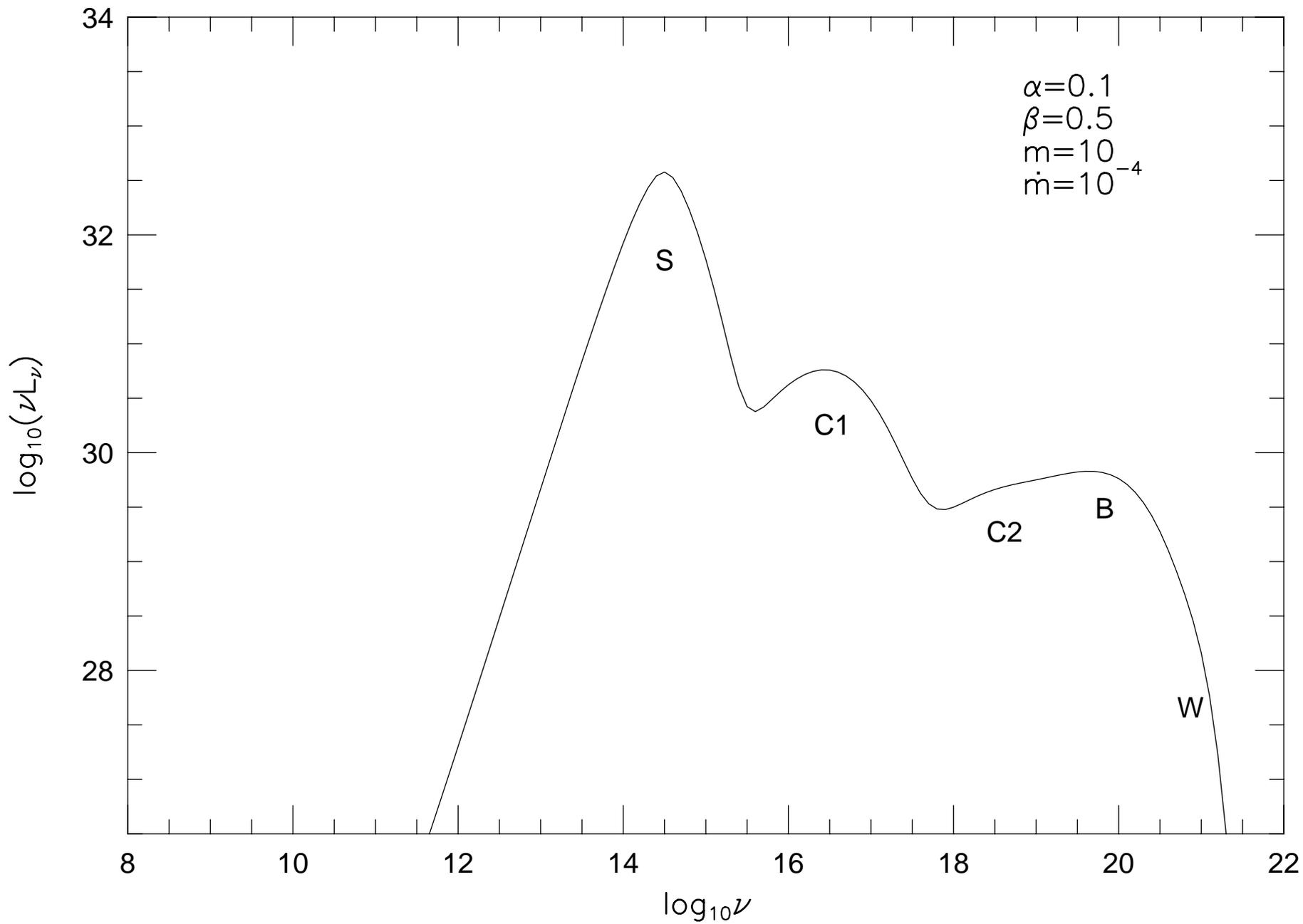
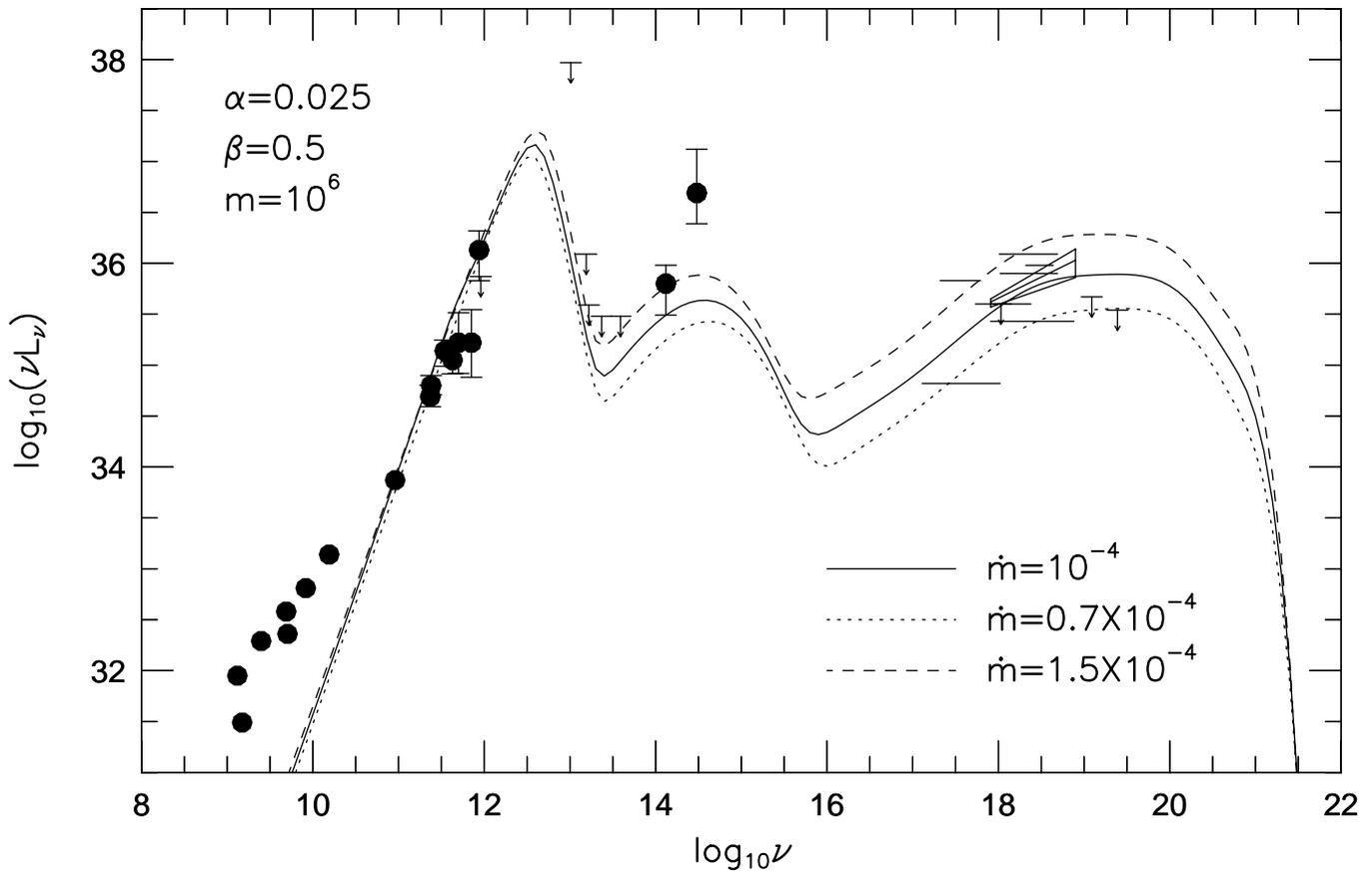
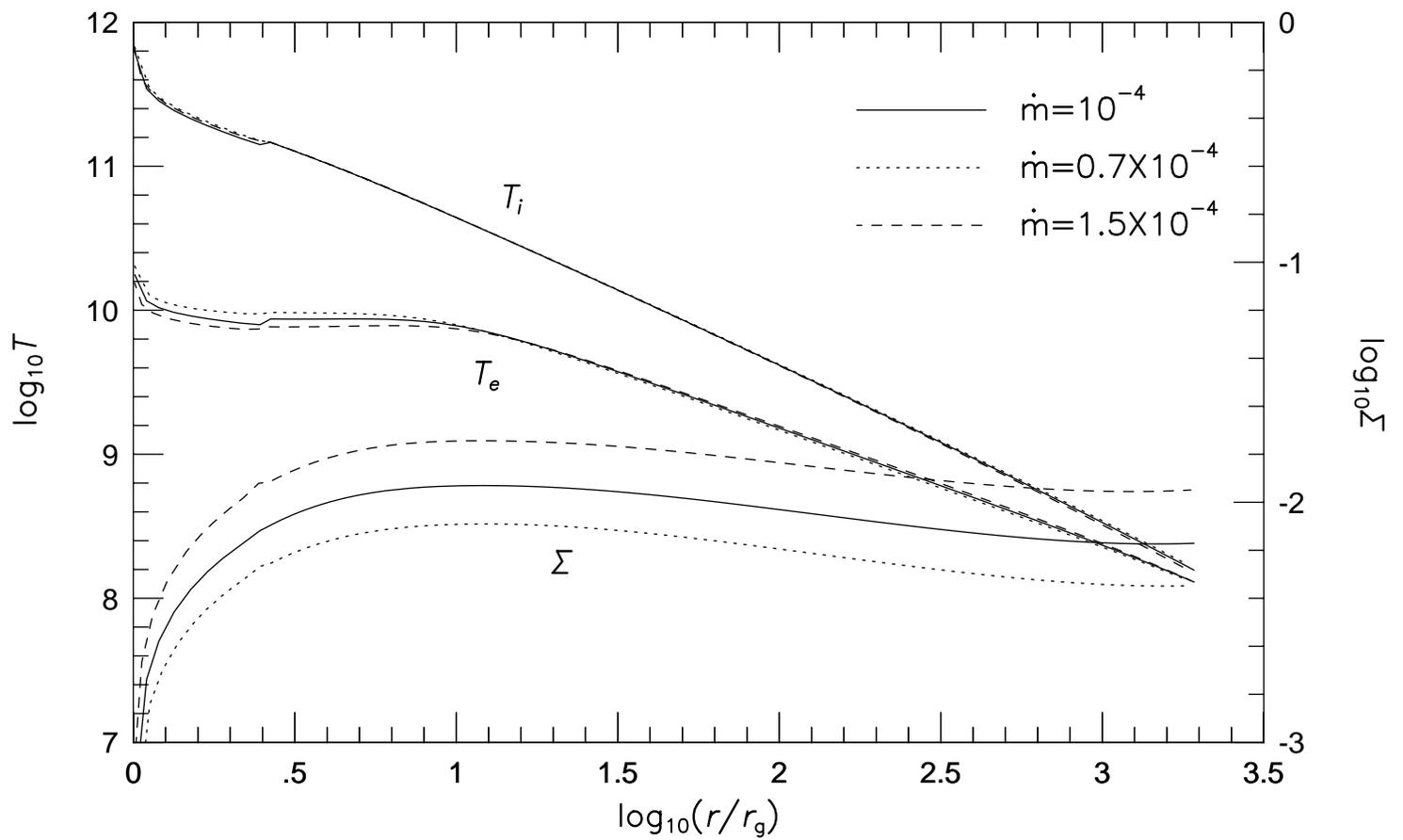
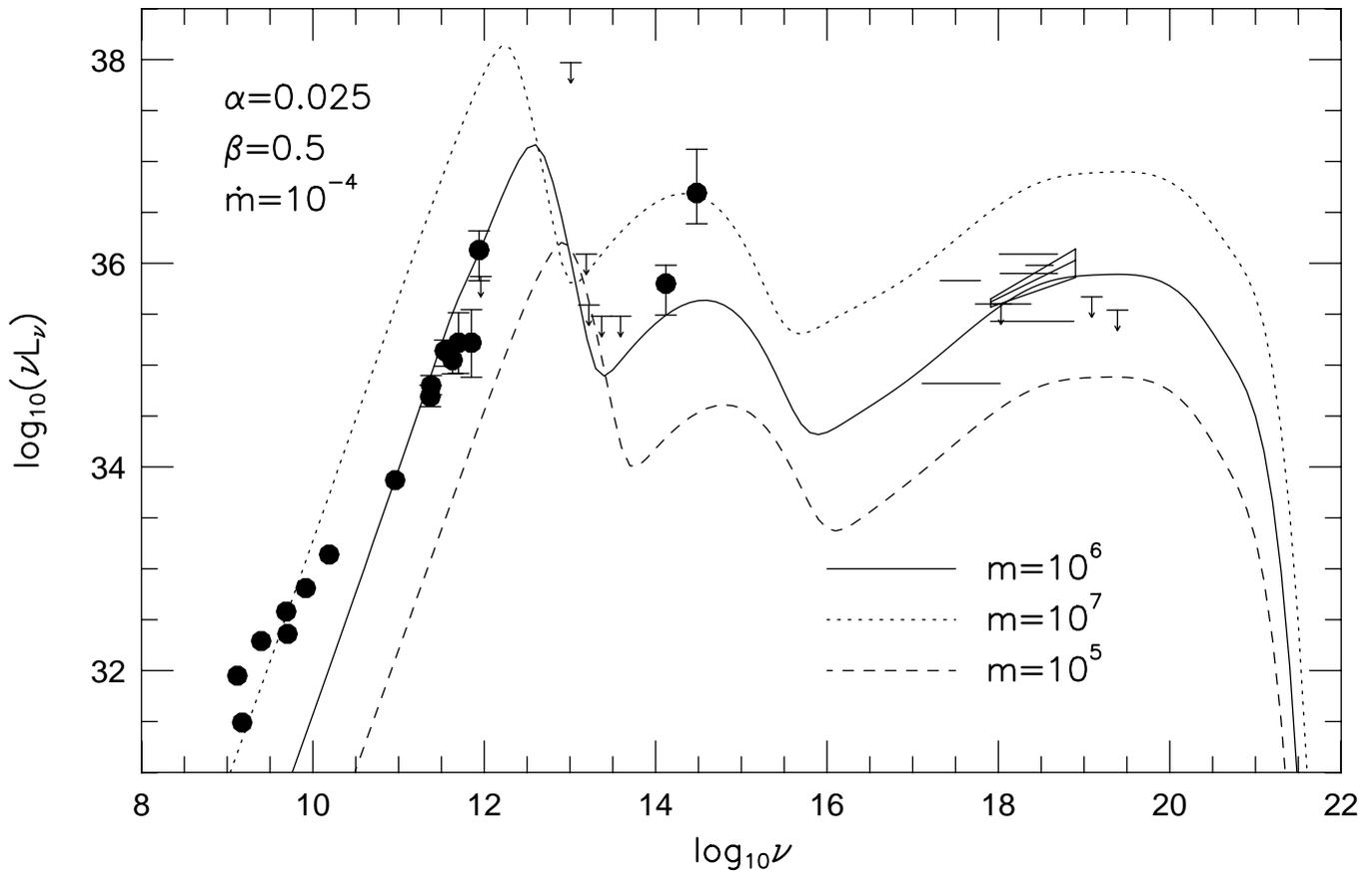
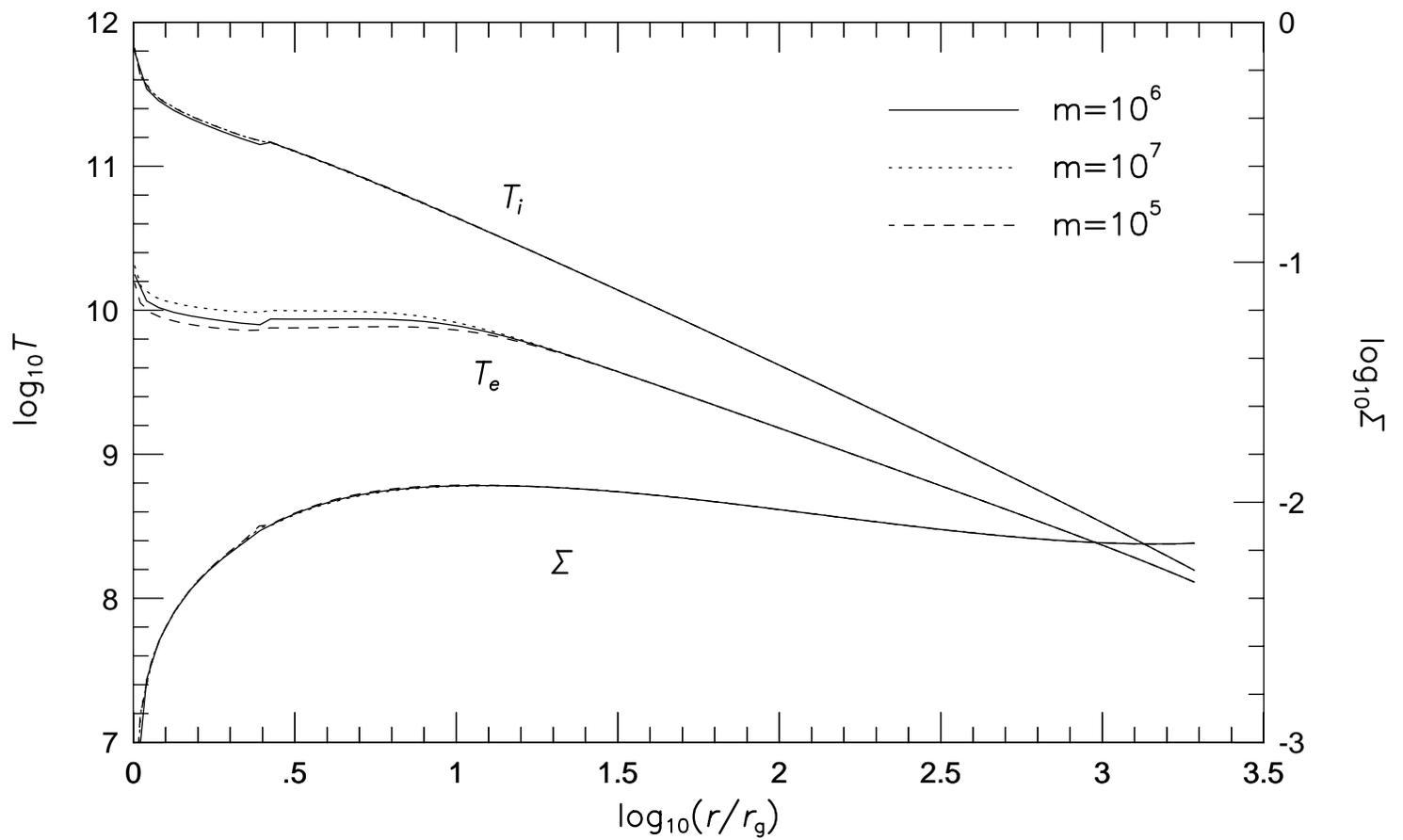
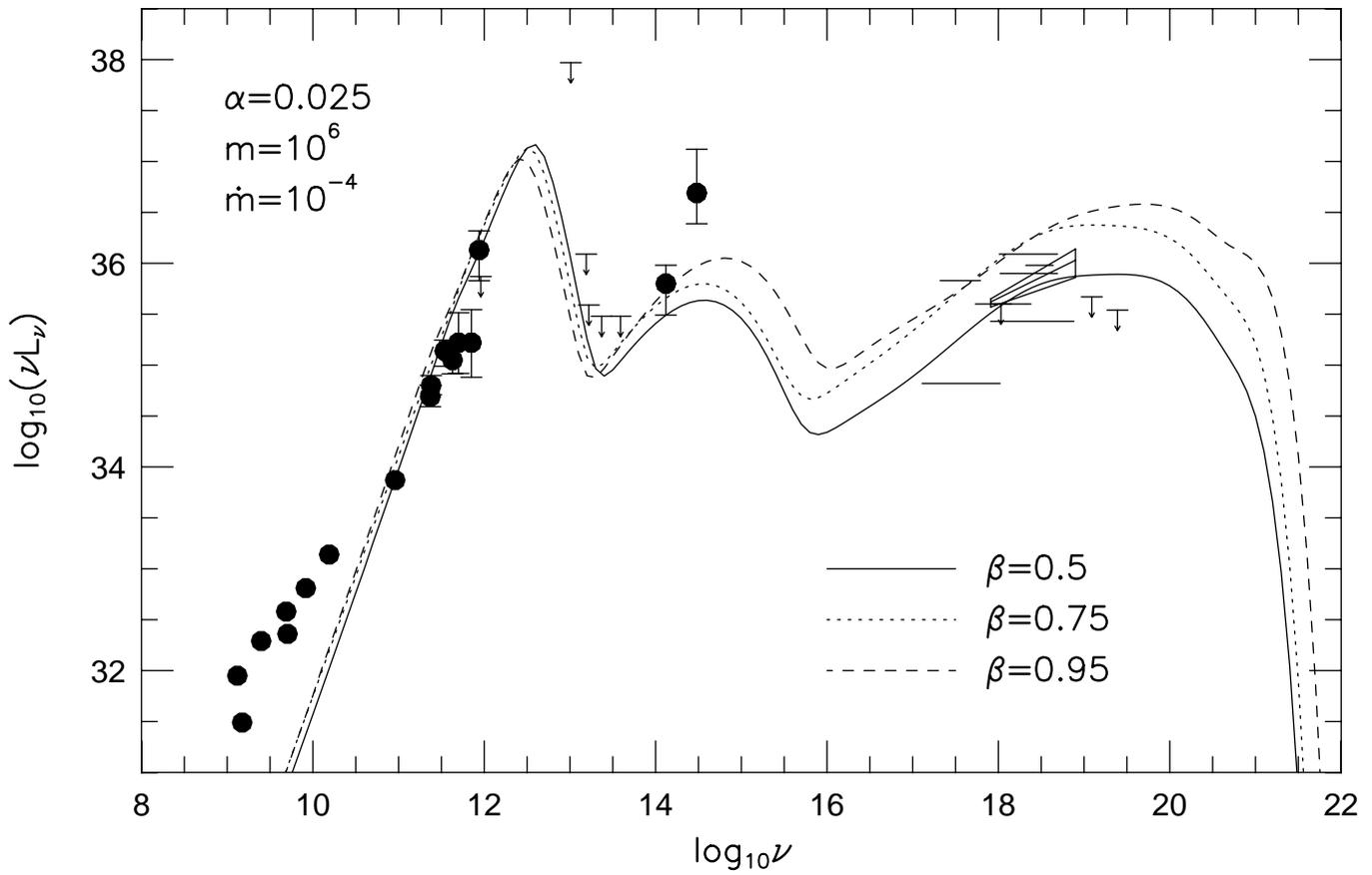
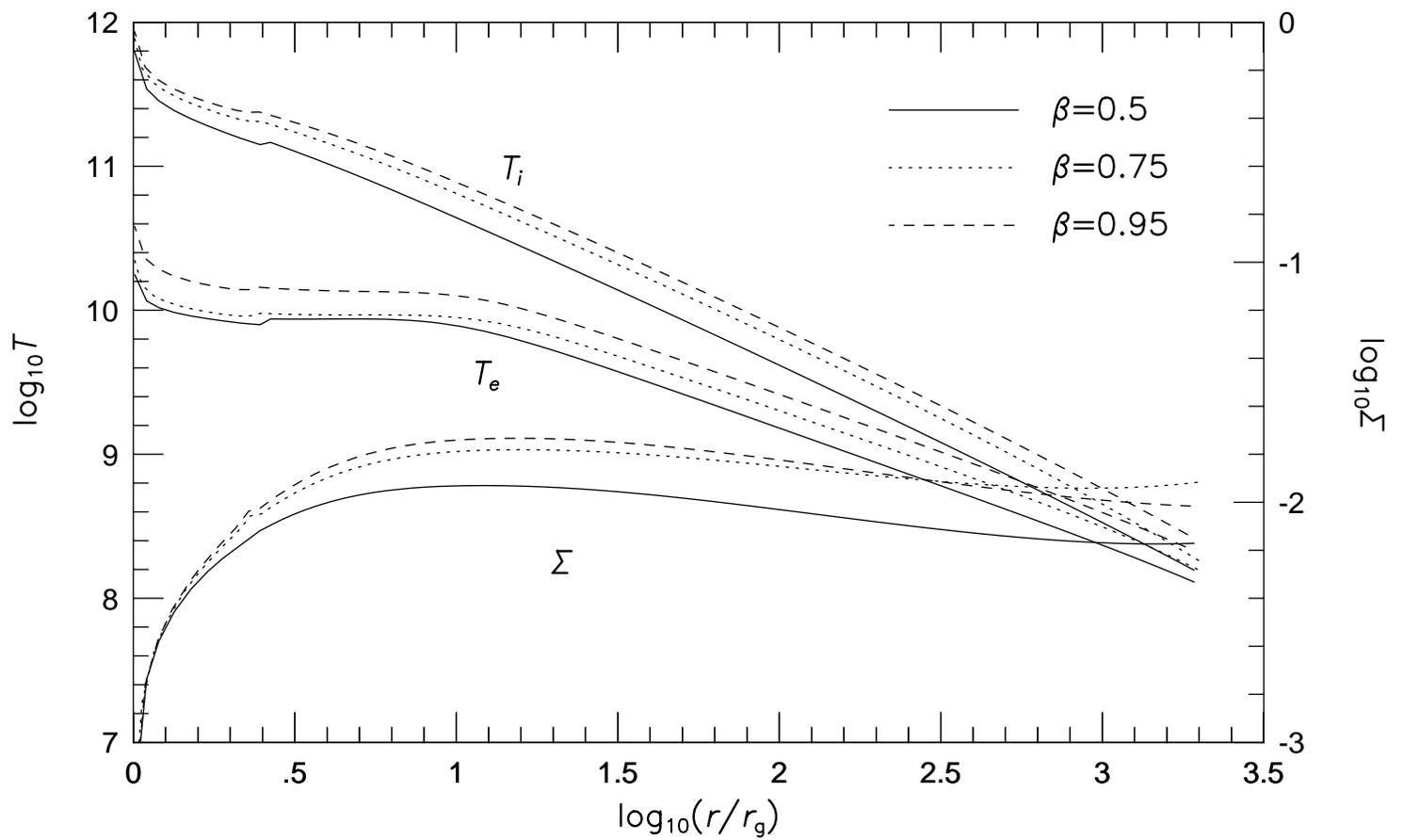
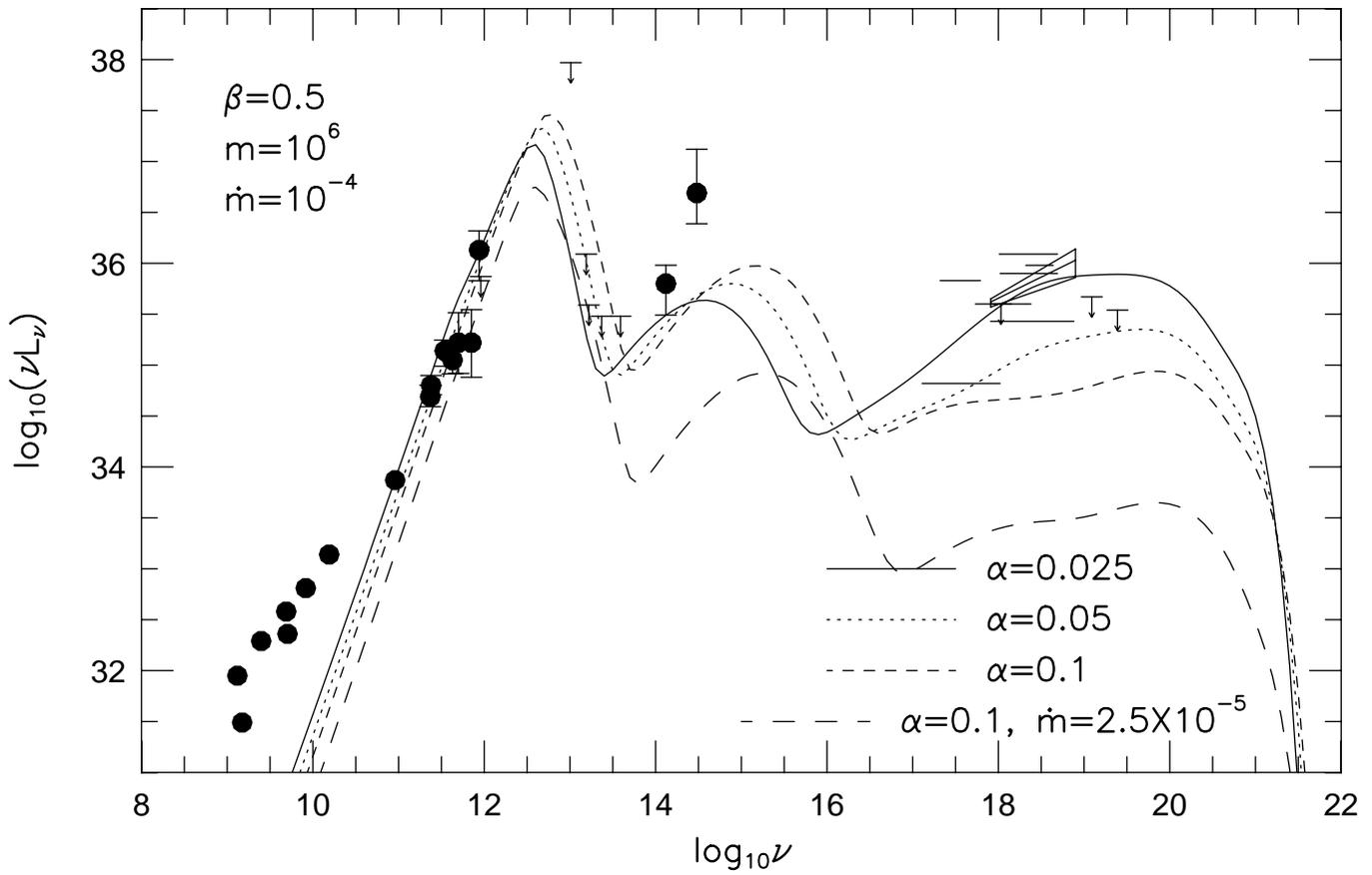
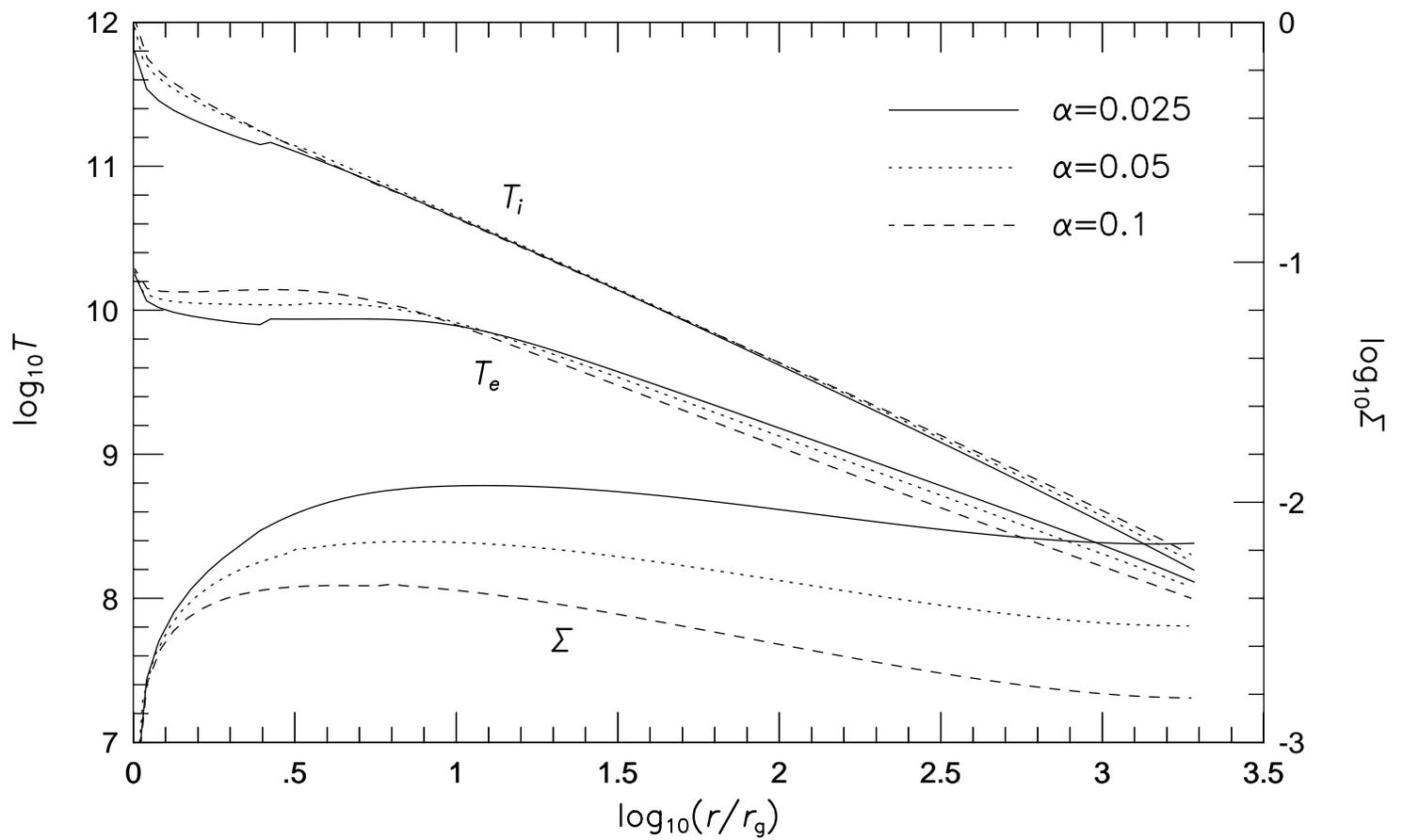
\end{document}